\documentclass[two column, times, trackchanges]{aastex62}
\usepackage{CJK}
\usepackage{color}
\usepackage{xcolor}
\usepackage{amsmath}
\usepackage{amstext}
\usepackage{amssymb}
\usepackage{enumitem}
\usepackage{mathtools}
\usepackage{newtxmath} 
\usepackage{hyperref}
\usepackage{textcomp}
\usepackage{gensymb}
\usepackage{amsmath}
\graphicspath{{./}{figures/}}

\received{September 20, 2019}
\revised{December 6, 2019}
\accepted{}
\submitjournal{ApJ}

\shorttitle{Debris Disks in Multi-Planet Systems}
\shortauthors{Dong et al.}
\begin{document}
\title{Debris Disks in Multi-Planet Systems: Are Our Inferences Compromised by Unseen Planets?}
\correspondingauthor{Jiayin Dong}
\email{jdong@psu.edu}

\author[0000-0002-3610-6953]{Jiayin Dong}
\affiliation{Department of Astronomy \& Astrophysics, The Pennsylvania State University, University Park, PA 16802, USA}
\affiliation{Center for Exoplanets \& Habitable Worlds, 525 Davey Laboratory, The Pennsylvania State University, University Park, PA 16802, USA}

\author[0000-0001-9677-1296]{Rebekah I. Dawson}
\affiliation{Department of Astronomy \& Astrophysics, The Pennsylvania State University, University Park, PA 16802, USA}
\affiliation{Center for Exoplanets \& Habitable Worlds, 525 Davey Laboratory, The Pennsylvania State University, University Park, PA 16802, USA}

\author[0000-0002-0711-4516]{Andrew Shannon}
\affiliation{LESIA, Observatoire de Paris, Universit\'{e} PSL, CNRS, Sorbonne Universit\'{e}, Universit\'{e} de Paris, 5 place Jules Janssen, 92195 Meudon, France}
\affiliation{Department of Astronomy \& Astrophysics, The Pennsylvania State University, University Park, PA 16802, USA}
\affiliation{Center for Exoplanets \& Habitable Worlds, 525 Davey Laboratory, The Pennsylvania State University, University Park, PA 16802, USA}
\author[0000-0002-2432-833X]{Sarah Morrison}
\affiliation{Department of Physics, Astronomy \& Materials Science, Missouri State University, Springfield, MO 65897, USA}
\affiliation{Department of Astronomy \& Astrophysics, The Pennsylvania State University, University Park, PA 16802, USA}
\affiliation{Center for Exoplanets \& Habitable Worlds, 525 Davey Laboratory, The Pennsylvania State University, University Park, PA 16802, USA}

\begin{abstract}
Resolved debris disk features (e.g., warps, offsets, edges and gaps, azimuthal asymmetries, radially thickened rings, scale heights) contain valuable information about the underlying planetary systems, such as the posited planet's mass, semi-major axis, and other orbital parameters. Most existing models assume a single planet is sculpting the disk feature, but recent observations of mature planetary systems (e.g., by radial velocity surveys or \textit{Kepler}) have revealed that many planets reside in multi-planet systems. Here we investigate if/how planet properties inferred from single-planet models are compromised when multiple planets reside in the system. For each disk feature, we build a two-planet model that includes a planet b with fixed parameters and a planet c with a full range of possible parameters. We investigate these two-planet systems and summarize the configurations for which assuming a single planet (i.e., planet b) leads to significantly flawed inferences of that planet's properties. We find that although disk features are usually primarily dominated by a single planet, when using single-planet models we are at risk of misinterpreting planet properties by orders of magnitude in extreme cases. Specifically, we are at high risk of misinterpreting planet properties from disk warps; at moderate risk from disk edges and gaps, radially thickened rings, and scale height features; and at low risk from host star-disk center offsets and azimuthal asymmetries. We summarize situations where we can infer the need to use a multi-planet model instead of a single-planet one from disk morphology dissimilarities.
\end{abstract}


\section{Introduction} \label{sec:intro}
Thousands of exoplanets have been discovered over the past few decades, including many by the high sensitivity \textit{Kepler} Mission \citep{boru10} and high precision radial velocity (RV) detection \citep[e.g.,][]{cumm08, howa10, mayo11}. Exoplanets observed with a wide variety of orbital and physical properties \citep[e.g., super-Earths, sub-Neptunes, hot/warm Jupiters;][and references therein]{winn15} require us to reassess the classical theory of planetary system formation and evolution built on the Solar System prototype. Most known exoplanets reside in systems that are a few Gyr old: they are the end states of a system's early formation and evolution. Planetary system formation theories \citep[e.g.][and references therein]{chia09, schl14, daws18} have been established based on these end states to explain the origins of planetary systems. One of the most direct ways to test these theories and answer the fundamental questions of planetary system formation and evolution (e.g., where do planets form and how do they migrate?) is to characterize young planetary systems during their early states of formation and evolution \citep{su19}.

\begin{figure*}
\plotone{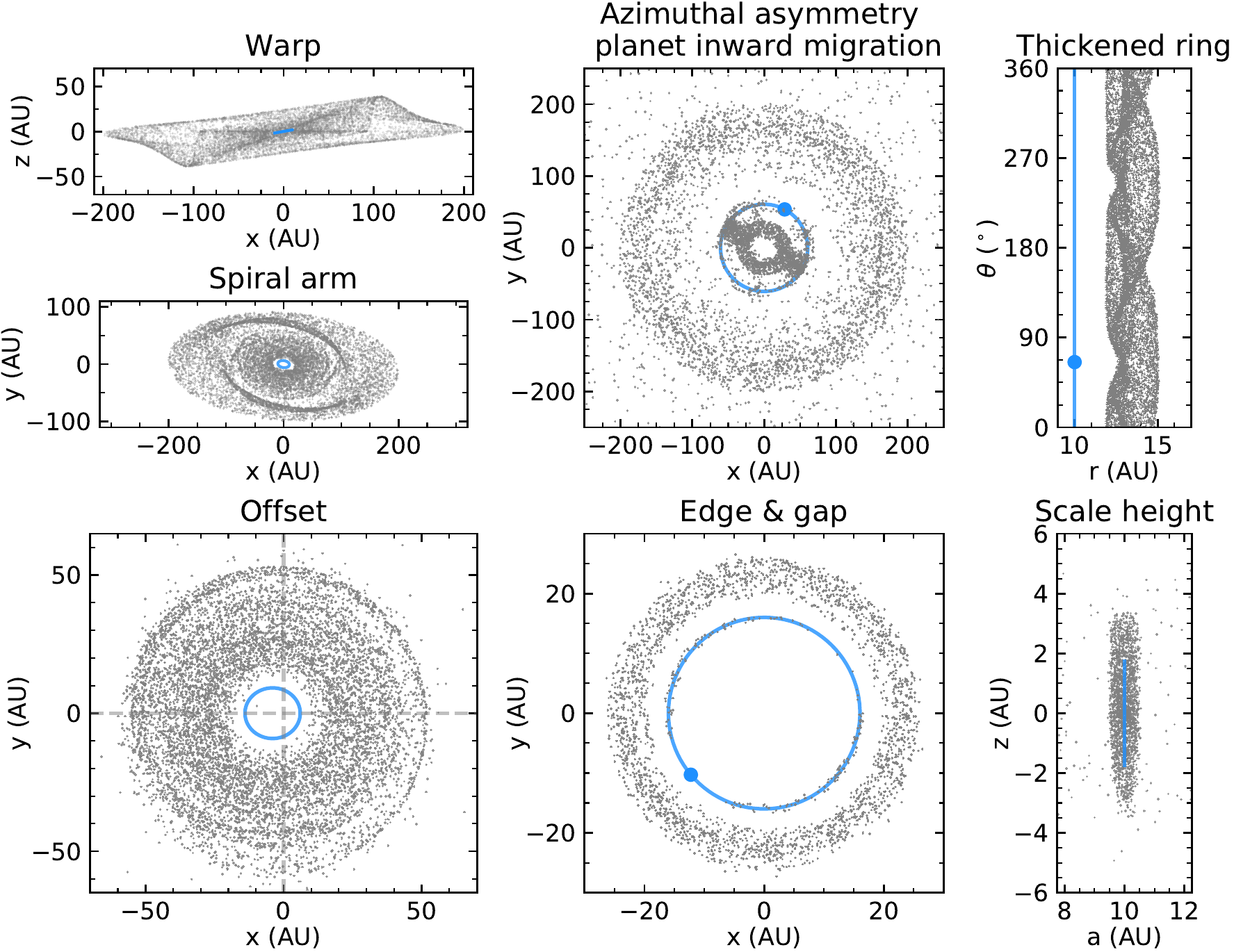}
\caption{A gallery of debris disk features sculpted by a single planet simulated with the $N$-body integrator \texttt{REBOUND} \citep{rein15a, rein15b} and \texttt{REBOUNDx} \citep{tama19}. Planets and their orbits are labeled in blue as dots and lines. Planetesimals (i.e., simulated as massless test particles) are labeled in grey. Planet locations are only indicated in systems where disk morphologies depend on their instantaneous locations. The left, middle, and right columns are corresponding to secular, resonant, and synodic features.} \label{fig:summary}
\end{figure*}

It is challenging to apply the most prolific exoplanet detection techniques to young planetary systems. The strong stellar activity noise \citep[e.g., flares, flicker, and jitter;][]{bast14} of young host stars can easily suppress or mimic exoplanet signals in transit and radial velocity observations. Although direct imaging detection (e.g., Gemini Planet Imager, or GPI; Spectro-Polarimetric High-contrast Exoplanet REsearch instrument, or SPHERE) is most feasible for young planetary systems, it is challenging as well because the high planet-star flux contrast results in low sensitivity to low mass planets. However, we can still probe young planetary systems indirectly through their debris disks. A debris disk, a circumstellar disk of dust and planetesimals orbiting its host star, has been detected around roughly 17--20\% solar-type stars \citep[][and references therein]{sibt18, hugh18}. The dust in a debris disk is continuously produced through collisions of planetesimals that are gravitationally redistributed by the posited underlying planets \citep{wyat08}. By investigating debris disk features (here we focus on warps, offsets, edges and gaps, azimuthal asymmetries, radially thickened rings, scale heights) induced by planet-disk interactions, we can characterize the corresponding planetary system properties, such as the planet mass, its orbital parameters, and the timescale expected for a disk to evolve the observed features \citep[see][and references therein]{hugh18}. Debris disk features have been well studied in previous literature using single-planet models. As shown in Figure~\ref{fig:summary}, all of the following disk features may be interpreted as perturbations from a single planet.\footnote{Figure~\ref{fig:summary} simulation setup can be found in Appendix \ref{setup}.}

\begin{itemize}[leftmargin=*]
    \setlength\itemsep{-0.2em}
    \item Warps: A warped disk has an inner disk inclined with respective to the outer disk. The disk needs to almost be edge-on so that the warp will be observed. A warp has often been interpreted as the signpost of a planet on an inclined orbit \citep[e.g., $\beta$-Pictoris;][]{moui97, auge01}.
    \item Spiral arms: The spiral arm feature refers to a disk with two symmetric spiral arms \citep[e.g., HD 141569;\footnote{HD 141569 could be a ``hybrid" disk between the stage of protoplanetary and debris disks \citep{mile18}. If so, the spiral arms might be a hydrodynamic effect.}][]{bill15, koni16}. The feature can be explained as the projected view of a warped disk and thus serves as the signpost of an inclined planet.
    \item Offsets: An offset disk has an evident geometric center offset from its central star. This feature has been interpreted as the signpost of a planet on an eccentric orbit \citep[e.g., Fomalhaut, HR 4796A;][]{kala05, macg17, tele00}. \citet{lee16} demonstrated an offset ring could present various morphologies (e.g., rings, needles, moths) in scattered light images caused by different viewing angles.
    \item Edges \& gaps: The edge and gap feature refers to a disk with gaps and sharp edges \citep[e.g., HR 8799;][]{su09, matt14, wiln18}. A void of planetesimals across a certain region of a disk indicates a sculptor with significant mass \citep[i.e., a planet clearing its chaotic zone;][]{wisd80}.
    \item Azimuthal asymmetries: Clumpy features observed in a disk \citep[e.g., $\beta$-Pictoris;][]{dent14} may indicate the existence of a migrating planet that captured and redistributed planetesimals to certain longitudes \citep{wyat03, rech08, must11}. \citet{shan15} demonstrated small dust grains migrating under Poynting-Robertson drag may also get trapped in resonance with planets and produce clumpy structures in scattered light images. 
    \item Radially thickened rings: A radially thickened ring (e.g., Fomalhaut, HR 4796A) has often been interpreted as a planet close to the inner edge of a debris ring stirring up nearby material \citep{chia09, rodi14}.
    \item Scale heights: A substantial disk scale height \citep[e.g., AU Microscopii, or AU Mic;][]{kris05, macg13, dale19} has been interpreted as perturbation of a large stirring body in the planetary system \citep{quil07, theb07}.
\end{itemize}

These resolved debris disk features contain valuable information about the underlying planetary systems, such as the sculpting planet's mass and orbital parameters. Most existing models assume that a single planet is sculpting the disk feature. However, recent observations of mature planetary systems (e.g., by radial velocity surveys and \textit{Kepler}) have revealed that many planets reside in multi-planet systems \citep[e.g., see][and references therein]{winn15}. In this paper, we investigate if/how planet properties inferred from single-planet models are compromised when multiple planets reside in the system. We categorize disk features listed in Figure~\ref{fig:summary} into three categories: secular, resonant, and synodic, according to their physical processes and timescales. Warps, spiral arms, and offsets are secular features driven by the interchanging angular momentum of planets and planetesimals over a long timescale (Section~\ref{sec:secular}). Edges, gaps, and azimuthal asymmetries are resonant features that result from planets clearing nearby regions due to overlapping resonances or capturing planetesimals into orbital resonances (Section~\ref{sec:resonant}). Radially thickened rings and scale heights are synodic features that occur when planets stir up nearby material either radially or vertically over a short timescale (Section~\ref{sec:synodic}). For each disk feature, we build two-planet models that include a planet b with fixed parameters and a planet c with a full range of possible parameters. We characterize planet properties from the disk features in these two-planet systems and summarize the configurations for which assuming a single planet (i.e., planet b) leads to significantly flawed inferences of that planet's properties. Our model considers planetesimals as massless, collisionless, and radiation-free test particles and studies how the gravitational influence of planets affects their distributions. Detailed modeling of other physical processes, such as collisions between planetesimals, is deferred to future work. We discuss secular features (warps, offsets), resonant features (edges and gaps, azimuthal asymmetries), and synodic features (radially thickened rings, scale heights) in Section~\ref{sec:secular}, \ref{sec:resonant}, and \ref{sec:synodic}, respectively. We summarize our main results and list situations where we can infer the need to use a multi-planet model instead of a single-planet one in Section \ref{sec:conclusion}.

\section{Secular Features in 2-planet Systems} \label{sec:secular}
Secular features are driven by the interchanging angular momentum of planets and planetesimals over a long timescale. Through secular interactions, a planet on an inclined or eccentric orbit modifies the disk morphology by changing the planetesimals' inclinations or eccentricities. Disk feature properties (e.g., a warped disk's warp location and inclination) can constrain the underlying posited planet's mass, semi-major axis, and other orbital parameters. In Section \ref{sec:secular}, we investigate if/how our inferences of these planet properties are compromised using a single-planet model when multiple planets reside in the system. To feasibly explore a wide range of two-planet configurations each with thousands of planetesimals, we take advantage of the Laplace-Lagrange secular theory, a first-order approximation in $e$ and $i$ of the secular terms in the disturbing function. The Laplace-Lagrange approximation is appropriate for  planets with low inclinations ($i\lesssim20\degree$) and eccentricities ($e\lesssim0.3$) (\citealt{murr99}, Chapter 7; see Appendix~\ref{secular code} in a nutshell). The analytical approximation, compared to $N$-body simulations, greatly reduces the computation time and allow us to explore hundreds of two-planet configurations in a reasonable time. 

\begin{figure*}
    \epsscale{1.}
    \plotone{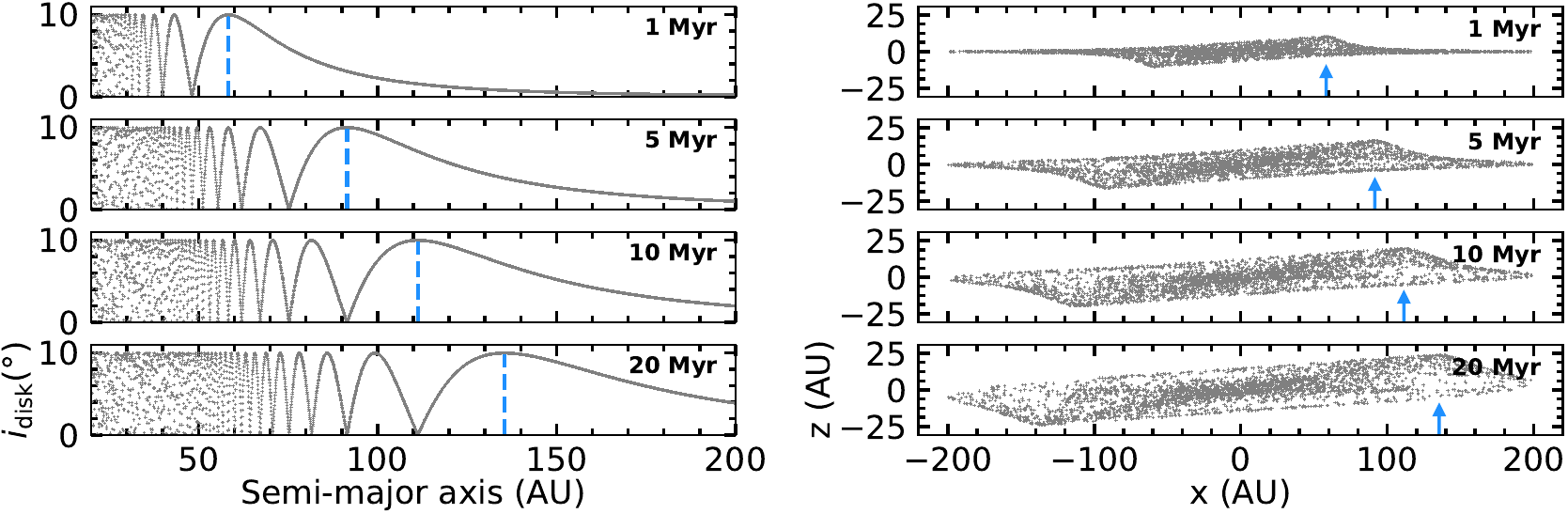}
    \caption{The evolution of a warp over time, demonstrated with planetesimal instantaneous inclination versus semi-major axis (left panel) and the projected image of the warp along the line of nodes of the planet (right panel). The warp is produced by a 10 Jupiter-mass planet with a semi-major axis of 10 AU and an inclination of 5$\degree$. The warp production time (i.e., the time taken to produce the warp) is labeled in each panel. Planetesimals' inclinations oscillate between zero and twice the planet's inclination at different nodal precession rates. The warp at a certain warp production time is located at the semi-major axis where planetesimals are just reaching to their peak inclination in their first secular oscillation cycle, labeled as the blue dashed line or arrow.}
    \label{fig:warp_evol}
\end{figure*}

\subsection{Warps}\label{sec:warp}
A warped disk, an inner disk inclined with respect to the outer disk, has been interpreted as evidence for a sculpting planet on an inclined orbit. The vertical warp gradually moves outward in the disk with a rate depending on the sculpting planet's mass and semi-major axis. Figure~\ref{fig:warp_evol} demonstrates how a warp moves outward in a disk as a function of the warp production time. The warp production time is the time to produce the observed warp (i.e., calculated from the system's age minus the planet formation time). We approximate the warp production time by assuming the timescale for the planet to form and get onto its current orbit is either short compared to the warp production time (i.e., the planet \emph{instantaneously} appears with its present day properties) and/or planet formation happened during gas disk stage when planetesimals were protected from planet by gas. With an estimation of the warp production time $\tau$ and the warp location $a_{\textrm{warp}}$, the planet's mass and semi-major axis ($m_{\textrm{b}}$ and $a_{\textrm{b}}$) can be constrained, although with degeneracy:
\begin{equation} \label{eqn:A}
    \frac{\pi}{\tau} = n_{\textrm{warp}}\frac{1}{4}\frac{m_{\textrm{b}}}{m_*}\alpha_{\textrm{b}}\bar{\alpha}_{\textrm{b}}b_{3/2}^{(1)}(\alpha_{\textrm{b}}),
\end{equation}
where $n_{\textrm{warp}}$ is the orbital frequency of a planetesimal at $a_{\textrm{warp}}$, the semi-major axis of the warp, $m_*$ is the host star mass, $b_{3/2}^{(1)}$ is the Laplace coefficient, and $\alpha_{\textrm{b}}$ and $\bar{\alpha}_{\textrm{b}}$ follow the formula:
\begin{equation} \label{eqn: alpha_b}
        \alpha_{\textrm{b}} \text{, } \bar{\alpha}_{\textrm{b}} = 
            \begin{dcases}                a_{\textrm{b}}/a_{\textrm{warp}} \text{, } 1 &\text{\quad if \quad} a_{\textrm{b}} < a_{\textrm{warp}} \\   a_{\textrm{warp}}/a_{\textrm{b}} &\text{\quad if \quad} a_{\textrm{b}} > a_{\textrm{warp}}.
            \end{dcases}
\end{equation}
Moreover, the planet's inclination can be inferred from either the overall disk inclination or the warp inclination:
\begin{equation} \label{eqn:i}
      i_{\textrm{b}} = i_{\textrm{disk}}.
\end{equation}
One of the best-studied warped disks is the debris disk of $\beta$-Pictoris ($\beta$-Pic). A vertical warp at approximately 85 au from the star is observed in optical and near-infrared wavelengths and has been interpreted as an outcome of a giant planet on an inclined orbit \citep{moui97, auge01}. \citet{lagr09, lagr10} later discovered $\beta$-Pic b, a directly imaged planet. \citet{daws11} further modeled $\beta$-Pic b, the posited planet generating the warp, using orbital constraints from the directly imaged planet and showed that planet b, rather than another planet, must be responsible for the warp. We discuss the tentative detection of $\beta$-Pic c \citep[$m_\textrm{c}$ = 9 M$_{\textrm{Jup}}$ and $a_\textrm{c}$ = 2.7 au;][]{lagr19} in the context of our findings at the end of Section~\ref{sec:warp_mass_err}.

\begin{figure*}
\epsscale{1.}
\plotone{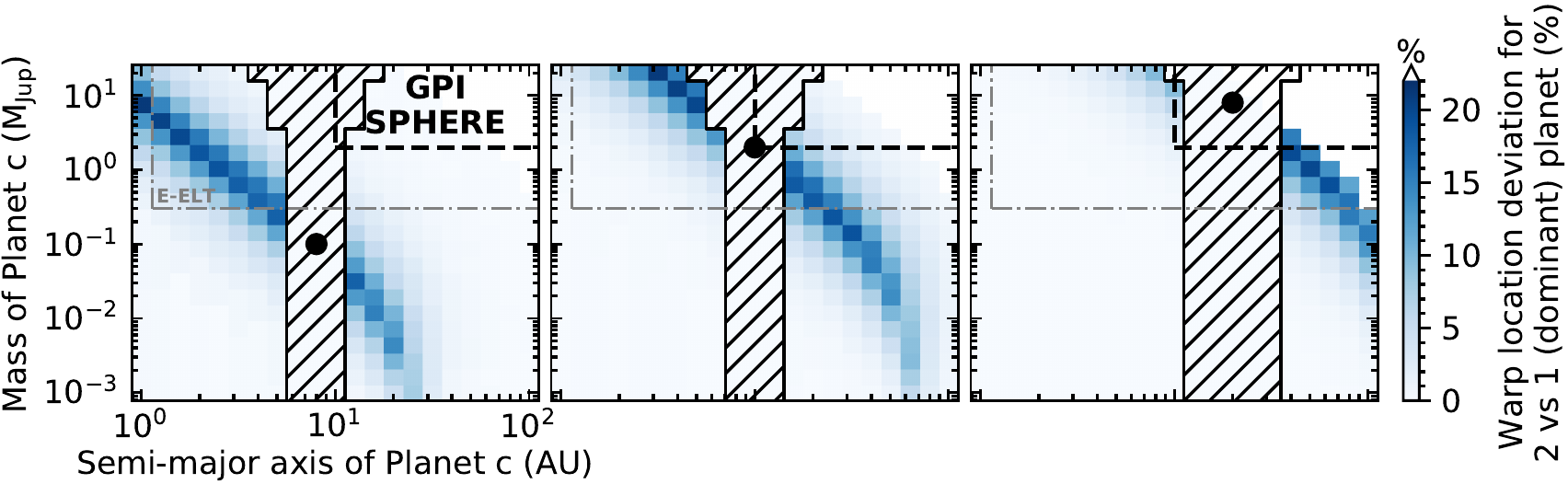} 
\caption{Warp location deviation in percentage for two-planet configurations versus one-dominant-planet configurations, where the dominant planet is defined as the planet that primarily sculpted the warp. The warp location deviation is calculated from Equation~(\ref{eqn:deviation}). Planet b, labeled as a black dot in each panel, has a mass $m_{\textrm{b}} = 0.1/2/8$ M$_{\textrm{Jup}}$ and a semi-major axis  $a_{\textrm{b}} = 8/10/20$ au. Planet c has a mass range of 0.001--20 M$_{\textrm{Jup}}$ and a semi-major axis range of 1--100 au. All systems have a warp production time of 10 Myr. The white region in the top right corner corresponds to systems for which warps move beyond 200 au (i.e., the edge of the disk) in 10 Myr. The hatch filled region corresponds to configurations that fail to satisfy the analytical dynamical stability criterion given in Equation~(\ref{eqn:stability}). The black dashed box in the top right corner represents roughly the parameter space GPI and SPHERE are sensitive to \citep[for a 30 Myr-old star at 30 pc;][]{bowl16}. The grey dashed box represents the predicted performance of E-ELT METIS L-band \citep[for a 30 Myr-old star at 30 pc;][]{quan15}. In each panel, planet b dominates the disk morphology in configurations below the dark blue strip (i.e., the bottom left corner), whereas planet c dominates beyond the strip (i.e., the upper right corner).}
\label{fig:warp_dev}
\end{figure*}

\begin{figure*}
\epsscale{1.}
\plotone{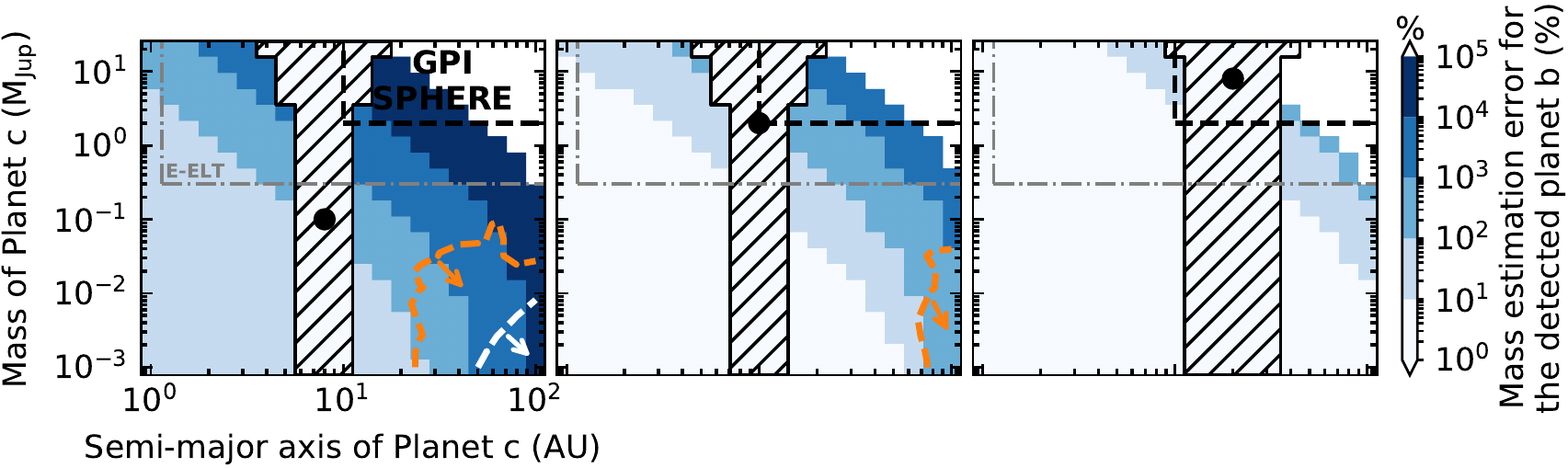}  
\caption{Mass estimation error in percentage for the detected planet b for two-planet configurations versus the planet b-only configuration, calculated from Equation~(\ref{eqn:warp_mass_error}). Planet b's mass and semi-major axis are indicated as the black dot in each panel. The white region in the upper right corner represents systems whose warps move beyond 200 au in 10 Myr. The hatched filled region is labeled for unstable configurations given in Equation~(\ref{eqn:stability}). The black dashed box in the top right corner represents roughly the parameter space GPI and SPHERE are sensitive to \citep[for a 30 Myr-old star at 30 pc;][]{bowl16}. The grey dashed box represents the predicted performance of E-ELT METIS L-band \citep[for a 30 Myr-old star at 30 pc;][]{quan15}. Colorscales indicate levels of mass estimation error. The darker the color, the more significant the error is caused by the undetected planet c. Configurations under the orange curve and the white curve correspond to the disk median inclination being reduced by 85\% and 65\%, respectively, compared to single-planet configurations. We categorize these configurations into two types of disk morphology dissimilarities between a single-planet system and a two-planet system, demonstrated in Figure~\ref{fig:warp_slope} and \ref{fig:warps} and explained in Section~\ref{sec:dissimilarity}.} \label{fig:mass_error}
\end{figure*}

\subsubsection{Mass estimation error using a single-planet model}\label{sec:warp_mass_err}
For systems like $\beta$-Pic with the detection of both a warped disk and a directly imaged planet, the planet's mass can be constrained from the warp location and the planet's semi-major axis. This approach can be applied to other warped-disk systems that will be discovered in future missions (e.g., James Webb Space Telescope, or \textit{JWST}; \citealt{beic19, chen19, bran19}; The Large UV/Optical/IR Surveyor, or \textit{LUVOIR}; \citealt{robe19, debe19}; The European Extremely Large Telescope, or E-ELT; \citealt{bran14, bran18}). However, our interpretation on the detected planet's mass could be compromised, if the system has one or more planets undetected. Here we investigate if/how our inference on planet b's mass is compromised if there is a hidden planet c, assuming planet b is the detected planet and planet c is the hidden planet. In our two-planet model, we fix planet b's parameters and explore a range of parameters for planet c (i.e., $m_\textrm{c}$ = 0.001--20 M$_{\textrm{Jup}}$ and $a_\textrm{c}$ = 1--100 au). Doing so allows us to explore a wide range of two-planet mass ratios and semi-major axis ratios. We first explore a simple case where planet b and c are coplanar ($i_\textrm{b}$ =  $i_\textrm{c}$). The forced inclination (i.e., the time-averaged inclination) of the planetesimal from such two planets can be written as $i_{\text{forced}} = i_\textrm{b,c}$ assuming all planetesimals begin with zero inclination and longitude of ascending node (i.e., a flat disk; see Appendix~\ref{secular code}). Planetesimals' inclinations oscillate between $0$--$2i_\textrm{b,c}$ at different nodal precession rates. Warps are located at semi-major axes where planetesimals are just reaching to their peak inclination in their first secular oscillation cycle. When multiple planets sculpt multiple warps, we characterize the warp location using the semi-major axis of the outermost warp since that warp will be the easiest one to observe.

We first study how far the warp location deviates in a two-planet system from a single-planet system. In other words, we study whether the warp location is mostly contributed from a single planet. For all simulations in this study, we assume a solar-mass host star. We consider a warp production time of 10 Myr, motivated by the earlier studies of the $\beta$-Pic warp \citep[e.g.,][]{daws11} that used a timescale of 10 Myr. To calculate the warp location deviation, we have:
\begin{equation} \label{eqn:deviation}
        \frac{a_{\textrm{warp}}' - \max(a_{\textrm{warp,b}}, a_{\textrm{warp,c}})}{\max(a_{\textrm{warp,b}}, a_{\textrm{warp,c}})} 100\%,
\end{equation}
where we compare the warp location sculpted by two planets, $a_{\textrm{warp}}'$, to the warp location sculpted by either planet b or planet c depending on which one has a greater value, $\max(a_{\textrm{warp,b}}, a_{\textrm{warp,c}})$. We summarize our results in Figure~\ref{fig:warp_dev}. The left, middle, and right panels of Figure~\ref{fig:warp_dev} present warp location deviations for three different planet b setups ($m_{\textrm{b}} = 0.1/2/8$ M$_{\textrm{Jup}}$ and $a_{\textrm{b}} = 8/10/20$ au). In the middle and right panels, it would be feasible to detect planet b by direct imaging using the GPI or SPHERE \citep{bowl16}. Planet c has a mass range of 0.001--20 M$_{\textrm{Jup}}$ and a semi-major axis range of 1--100 au. Dynamically unstable two-planet configurations are labeled as the hatched filled region using the criterion given in \citet{petr15}:
\begin{multline}\label{eqn:stability}
    a_\textrm{out}(1-e_\textrm{out})/[a_\textrm{in} (1+e_\textrm{in})] > \\
    2.4 \big[\textrm{max}(\mu_\textrm{in},\mu _\textrm{out})\big]^{1/3} (a_\textrm{out}/a_\textrm{in})^{1/2} +1.15,
\end{multline}
where $a_\textrm{out}$, $e_\textrm{out}$, and $\mu_\textrm{out}$ are the semi-major axis, eccentricity, and planet-star mass ratio for the outer planet, respectively, and $a_\textrm{in}$, $e_\textrm{in}$, and $\mu_\textrm{in}$ are for the inner planet. As shown in Figure~\ref{fig:warp_dev}, the warp location is dominated by a single planet in most two-planet configurations (i.e., configurations in light blue). Planet b dominates the disk morphology in configurations below the dark blue strip (i.e., the bottom left corner), whereas planet c dominates beyond the strip (i.e., the upper right corner). Which planet dominates the disk purely depends on the planets' contribution on the nodal precession rate. When planet b and planet c have a similar contribution, the warp location deviation is the highest and can be at most 22\%.\footnote{The maximum warp location deviation can be derived from setting the same nodal precession rate in a single-planet system and a two-planet system and solving for the warp location ratio.}

We also notice that in systems with one or two massive planets with large semi-major axes, the warp sculpted by it/them can easily move beyond the observable outer edge of the debris disk (i.e., $\sim$200 au) in a short timescale (i.e., $<$10 Myr). This region of parameter space is colored white in Figure~\ref{fig:warp_dev} and subsequent figures. 

On the one hand, Figure~\ref{fig:warp_dev} demonstrates in most configurations one planet dominates the warp location, but on the other hand, it could be problematic when the dominating planet is not the one directly imaged. The mass estimation for the detected planet b could be significantly flawed, if the warp is majorly sculpted by a hidden planet c. To characterize the mass estimation error in situations like this, we estimate planet b's mass from the warp location sculpted by two planets and compare the estimated mass to planet b's true mass. The mass estimation error is
\begin{equation} \label{eqn:warp_mass_error}
        \frac{ m_{\textrm{b}}'-m_{\textrm{b}}}{m_{\textrm{b}}}100\%,
\end{equation}
where $m_{\textrm{b}}'$ is the estimated mass of planet b and $m_{\textrm{b}}$ is planet b's true mass. As shown in Figure~\ref{fig:mass_error}, when planet c gradually dominates the disk morphology, the mass estimation error for planet b grows because we credit the disk feature to the wrong planet. The upper right black box in each panel indicates the parameter space to which GPI and SPHERE are sensitive \citep[assuming a 30 Myr-old star at 30 pc;][]{bowl16}; within that box, planet c is detectable around \emph{some} stars. The grey box represents the predicted parameter space E-ELT METIS L-band is sensitive to.\footnote{We estimate the E-ELT performance with an inner working angle of 0.038$\arcsec$ and a planet-star flux contrast of $10^{-8}$ for a 30 Myr-old star at 30 pc \citep{quan15}. The planet luminosity is estimated using Eqn. (1) in \cite{bowl16}.} The mass estimation error for planet b could be huge if planet c is either massive or close to the warp but not detected. The left panel of Figure~\ref{fig:mass_error} ($m_{\textrm{b}} = 0.1$ M$_{\textrm{Jup}}$, $a_{\textrm{b}}$ = 8 au) emphasizes two-planet configurations for which planet b's estimated mass could deviate significantly from its true mass. As shown in dark blue pixels, a large semi-major axis but low mass planet c may cause a mass estimation error as high as 10,000--100,000\% for planet b. The reason is that the observed warp is predominantly contributed by planet c with a large semi-major axis (i.e, $>$100 au). Due to planet c's low mass, the warp has not moved beyond of the outer edge of a debris disk yet. Attributing a warp so far from the star to a small semi-major axis planet b results in a huge mass estimation error for the planet. In contrast, if a massive planet b with a large semi-major axis is directly imaged (i.e., the right panel in Figure~\ref{fig:mass_error}), the mass estimation for the planet will be close to its true mass. The small estimation error in this case is primarily because a massive, large semi-major axis planet c is incompatible with a warp location inside the disk over 10 Myr.

We consider the tentative detection of $\beta$-Pic c \citep[$m_\textrm{c}$ = 9 M$_{\textrm{Jup}}$ and $a_\textrm{c}$ = 2.7 au;][]{lagr19} in the context of our findings. How would $\beta$-Pic c contribute to the observed warp? We can answer the question from Figure~\ref{fig:warp_dev}, the warp location deviation plot. $\beta$-Pic b's mass and semi-major axis ($m_\textrm{b}$ = 12 M$_{\textrm{Jup}}$ and $a_\textrm{b}$ = 8.9 au) are similar to the planet b setup in the right panel of Figure~\ref{fig:warp_dev} \citep{lagr12, bonn14}. Figure~\ref{fig:warp_dev} shows that the parameter space planet c lies in has minimal influence on the warp location. Therefore, $\beta$-Pic b remains as the dominant planet sculpting the observed warp. Furthermore, \citet{daws11}\footnote{We found that D+11 mistakenly used 1 solar-mass instead of 1.75 solar-mass in calculating the warp production time in their Figure 1. This error does not affect the conclusions of D+11.} demonstrated $\beta$-Pic b must be responsible for the warp, given the planet's inclination constraint from direct imaging and the observed disk inclination. A second planet is unlikely responsible for the warp because it would be hard for that planet to produce the warped disk morphology without exciting planet b's inclination. Our finding that a second planet could affect  the warp location in a system similar $\beta$-Pic (i.e., Figure~\ref{fig:warp_dev}) is compatible because we consider a more general case  where the planet's inclination is unknown.

\subsubsection{Disk morphology dissimilarities between a one-planet system and a two-planet system}\label{sec:dissimilarity}
\begin{figure}
\epsscale{1.}
\plotone{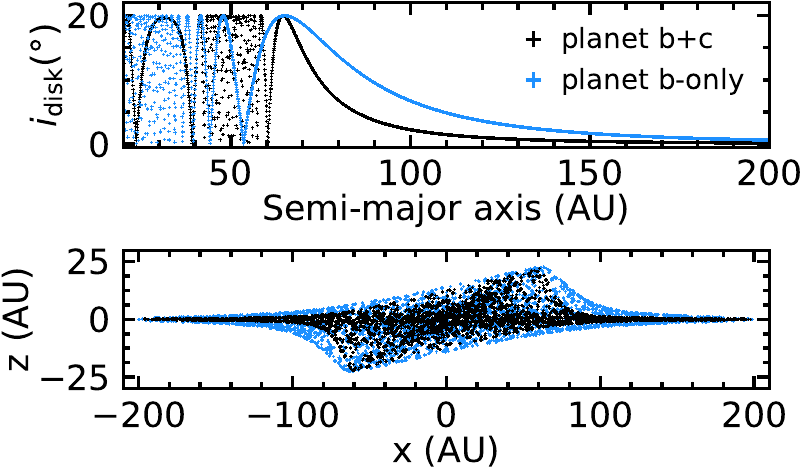}
\caption{Snapshot (10 Myr) of a disk's instantaneous inclination (upper panel) and morphology (lower panel) sculpted by planet b and c ($m_{\textrm{b}} = 0.1$ M$_{\textrm{Jup}}$, $a_{\textrm{b}}$ = 8 au, $m_{\textrm{c}} = 0.01$ M$_{\textrm{Jup}}$, $a_{\textrm{c}}$ = 50 au; in black) versus by planet b only ($m_{\textrm{b}}' = 2.35$ M$_{\textrm{Jup}}$, $a_{\textrm{b}}$ = 8 au; in blue). To emphasize the morphology differences, the axis aspect ratio is not unity (i.e., y:x is 3:2).} \label{fig:warp_slope}

\vspace*{\floatsep}

\plotone{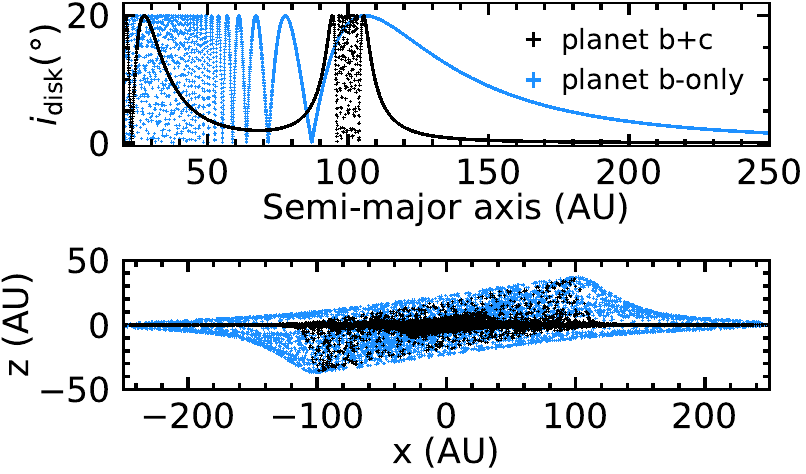}
\caption{Snapshot (10 Myr) of the disk's instantaneous inclination (upper panel) and morphology (lower panel) sculpted by planet b and c ($m_{\textrm{b}} = 0.1$ M$_{\textrm{Jup}}$, $a_{\textrm{b}}$ = 8 au, $m_{\textrm{c}} = 0.001$ M$_{\textrm{Jup}}$, $a_{\textrm{c}}$ = 100 au) versus by planet b only ($m_{\textrm{b}}' = 13.2$ M$_{\textrm{Jup}}$, $a_{\textrm{b}}$ = 8 au; in blue).}
\label{fig:warps}
\end{figure}

It is no surprise that the disk morphology sculpted by two planets will on some level differ from disk morphology sculpted by a single planet. If we are able to distinguish these differences, we may infer the need to use a multi-planet model instead of a single-planet one and thus avoid potential mass estimation errors. Here we summarize two types of disk morphology dissimilarities between a one-planet system and a two-planet system. The first is the thickness of the outer disk (i.e., how steeply the outer disk flattens from the warp), labeled as configurations within the orange dashed curve in Figure~\ref{fig:mass_error}. When the outer disk beyond the warp is primarily sculpted by a low mass planet c, the warp flattens steeply. Compared to a scenario where planet b is responsible for the warp, the outer disk is expected to be thinner, illustrated in Figure~\ref{fig:warp_slope}. As shown in Figure~\ref{fig:warp_slope}, the disk instantaneous inclination beyond the warp is much lower in a planet b and c-sculpting system (in black) than a planet b-sculpting system (in blue), even though both systems have the same warp location. Consequently, we observe a much sharper slope beyond the warp. Arrows in Figure~\ref{fig:mass_error} indicate the direction in which this feature becomes more obvious and observable (i.e., lower masses and larger semi-major axes for planet c). The second morphology difference between warps sculpted by one versus two planets corresponds to configurations under the white curve in Figure~\ref{fig:mass_error}. In these configurations, we expect to observe two warps sculpted by planet b and c individually, illustrated in Figure~\ref{fig:warps}. Given planet c's large semi-major axis and low mass, it sculpts an outer warp at around 105 au after 10 Myr evolution. Meanwhile, planet b sculpts a warp at around 28 au. The region between planet b and c remains flat and thus we observe two separate warps. For a given evolution timescale, this double warp feature is only present when planet b's mass and/or semi-major axis are sufficiently small, so we do not see this feature in the middle and right panels of Figure~\ref{fig:mass_error} (i.e., more massive planet b). It has a strong dependence on the warp production time, since after the inner warp and the outer warp meet, we will no longer observe multiple warps. 

\begin{figure}
\epsscale{1.}
\plotone{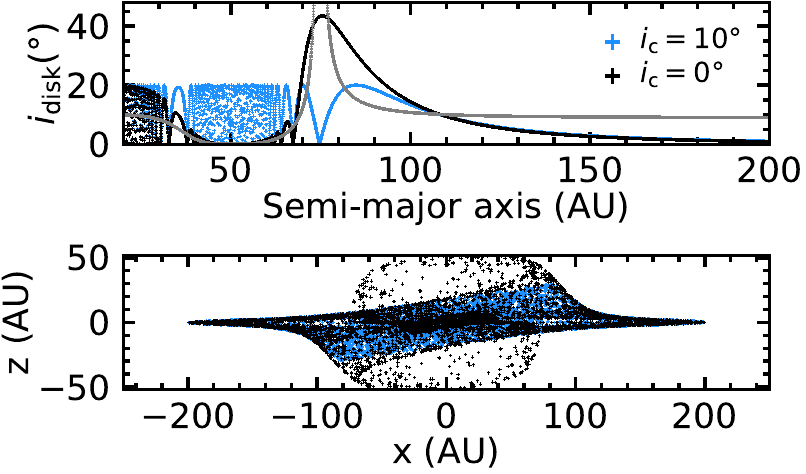}
\caption{Illustration of an inclination-type secular resonance modifying the location and height of a warp. Both simulations (i.e., blue and black) have an evolution timescale of 5.14 Myr (i.e., one secular cycle) and the same planet masses and semi-major axes: $m_{\textrm{b}} = 2$ M$_{\textrm{Jup}}$, $a_{\textrm{b}}$ = 10 au, $m_{\textrm{c}} = 0.1$ M$_{\textrm{Jup}}$, and $a_{\textrm{c}}$ = 50 au. In the blue system, $i_{\textrm{b}} = i_{\textrm{c}} = 10\degree$ (i.e., the planets are coplanar), whereas in the black system, $i_{\textrm{b}} = 10\degree$ and $i_{\textrm{c}} = 0\degree$ (i.e., the planets have a mutual inclination of 10 degrees). The upper panel presents the instantaneous inclinations of planetesimals at different semi-major axes. The grey curve is the forced inclination of the black system, with a secular resonance around 75 au that modifies the warp location. The lower panel is a projected view of disk morphology showing differences caused by the secular resonance.} \label{fig:sec_res}
\end{figure}

\begin{figure*}
\epsscale{1.}
\plotone{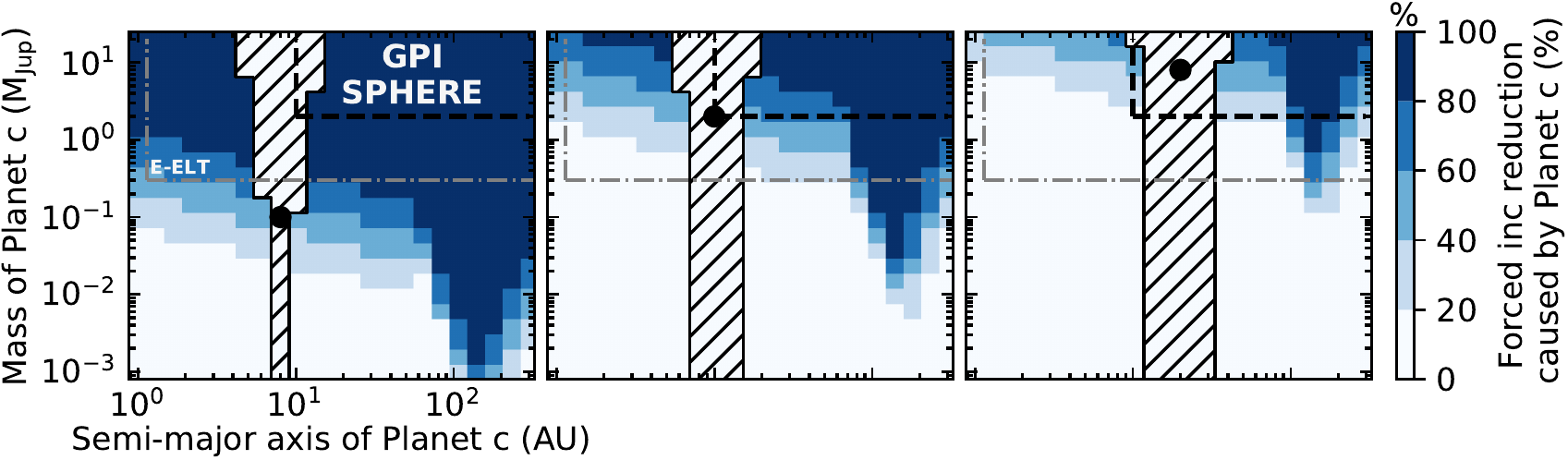}
\caption{Median disk forced inclination deviation in percentage for two-planet configurations versus planet b-only configuration, calculated from Equation~(\ref{eqn:dev}). Planet b, labeled as a black dot in each panel, has a mass of $m_{\textrm{b}} = 0.1/2/8$ M$_{\textrm{Jup}}$ and a semi-major axis of $a_{\textrm{b}} = 8/10/20$ au. Planet c has a mass range of 0.001--20 M$_{\textrm{Jup}}$ and a semi-major axis range of 1--300 au. Planet b has an inclination of 10$\degree$, whereas planet c has zero inclination. The hatched filled region corresponds to dynamically unstable configurations using the criterion given in Equation~(\ref{eqn:stability}). The black dashed box in the top right corner represents roughly the parameter space GPI and SPHERE are sensitive to \citep[for a 30 Myr-old star at 30 pc;][]{bowl16}. The grey dashed box represents the predicted performance of E-ELT METIS L-band \citep[for a 30 Myr-old star at 30 pc;][]{quan15}. The darkened region in each panel corresponds to planet c causing significant disk forced inclination deviations (i.e., $i_{\textrm{disk}}' \to 0\degree$).}
\label{fig:inc_err}
\end{figure*}

\begin{figure}
\epsscale{1.}
\plotone{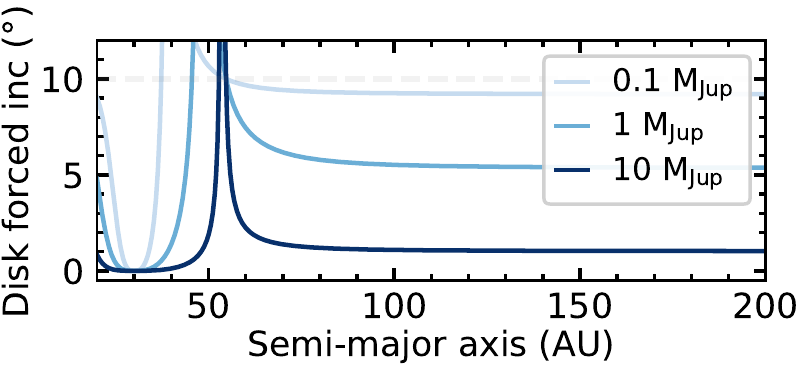}
\caption{Disk forced inclination as a function of the semi-major axis. In each system, planet b is a 2 Jupiter-mass planet with a semi-major axis of 10 au and a 10$\degree$ inclined orbit, the same as the middle panel of Figure~\ref{fig:inc_err}. Planet c has a semi-major axis of 30 au and zero inclination. Different curves represent disk forced inclination profiles in systems with different masses for planet c  ($m_{\textrm{c}}$ = 0.1, 1, 10 M$_{\textrm{Jup}}$). Increasing planet c's mass forces the disk to lower forced inclinations.} \label{fig:inc_layer}
\end{figure}

\begin{figure}
\epsscale{1.}
\plotone{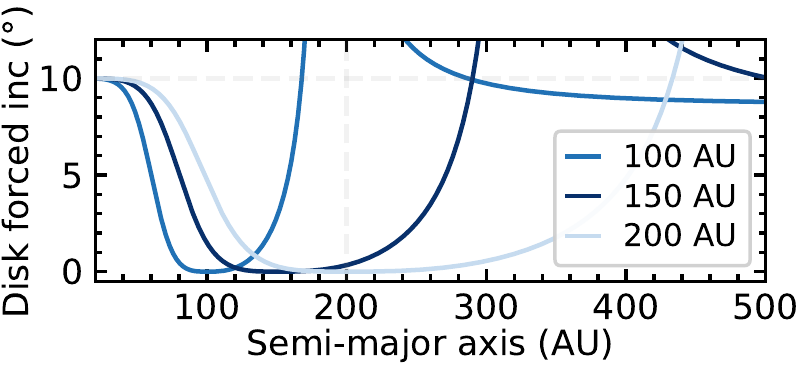}
\caption{Disk forced inclination as a function of the semi-major axis. In each system, planet b is a 2 Jupiter-mass planet with a semi-major axis of 10 au and a 10$\degree$ inclined orbit, the same as the middle panel of Figure~\ref{fig:inc_err}. Planet c has a mass of 0.1 M$_{\textrm{Jup}}$ and zero inclination. Different curves represent disk forced inclination profiles in systems with planet c having different semi-major axes ($a_{\textrm{c}}$ = 100, 150, 200 au). Systems in which planet c has a large semi-major axis are affected by a secular resonance (i.e., the spike around 180 au in $a_{\textrm{c}}$ = 100 au case).} \label{fig:inc_tri}
\end{figure}

\subsubsection{The effect of secular resonances}
The forced inclination of planetesimals sets the warp height and the overall disk inclination. We can understand the forced inclination as a time-averaged inclination on which planetesimals have their inclination oscillations centered. In systems where planet c has the same inclination as planet b, the forced inclination is a constant with  semi-major axis and is the same as the planets' inclination. However, in systems where planet c and planet b have different inclinations, the forced inclination is not a constant but depends on the planets' masses, semi-major axes, and inclinations. First, both planets will force nearby planetesimals to their inclined orbits and therefore the observed disk inclination will differ from a single-planet case. Secondly, secular resonances modify the disk forced inclination. When the nodal precession rates of planetesimals reach one of the characteristic frequencies of the planetary system, the planetesimals get excited to highly inclined orbits.\footnote{In Equation~(\ref{eqn:forced}), the forced inclination or eccentricity approaches infinity as $B \to f_i$ or $A \to g_i$ in the first-order secular approximation.} Usually the secular resonance has little effect on the morphology of a warped disk. Since we observe the disk edge-on, gaps cleared by resonances would not be easily observable. At certain warp production timescales, however, a warp could move near one of the secular resonance locations and the warp location could be modified by the resonance. In Figure~\ref{fig:sec_res}, we present a case where the secular resonance reduces the semi-major axis of a warp but increases the warp height. In this situation, we are at risk of moderately underestimating the planet's mass but overestimating the planet's inclination. Fortunately, a secular resonance can only modify the warp at a particular warp production time. Nevertheless, it would be helpful to take this possibility into consideration when characterizing planet properties. 

\begin{figure*}
\epsscale{1.}
\plotone{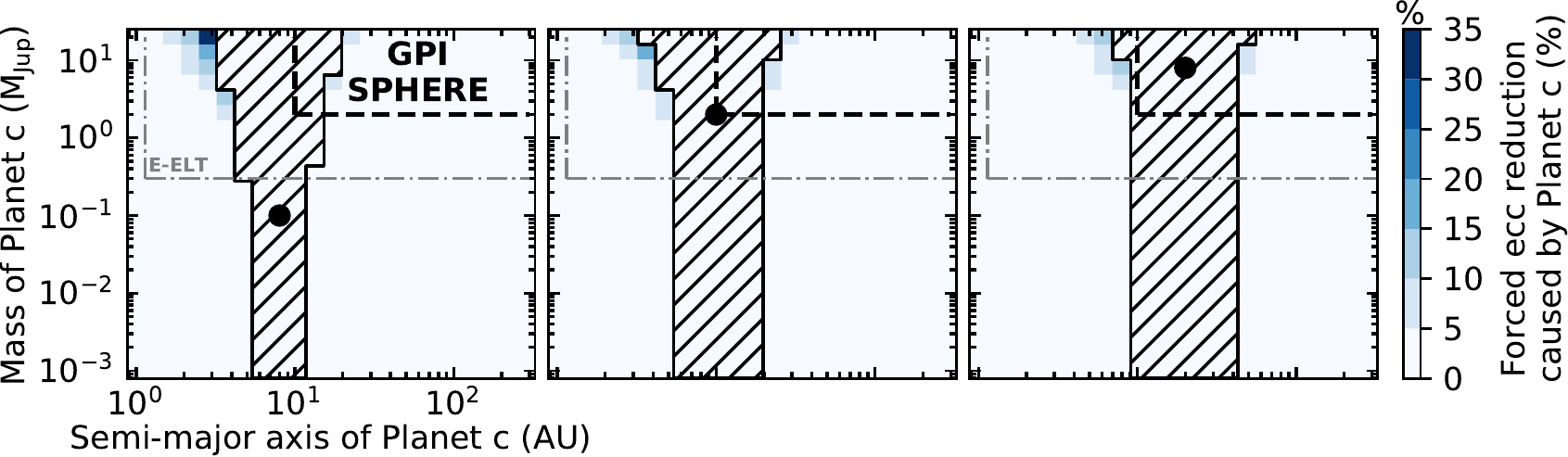}
\caption{Forced eccentricity reduction on a ring 2.5 Hill radii away from an eccentric orbit planet b ($a_{\textrm{ring}} = a_{\textrm{b}} + 2.5R_{\textrm{Hill,b}}$, $e_{\textrm{b}}$ = 0.2) caused by a zero eccentricity planet c ($e_{\textrm{c}}$ = 0). Planet b, labeled as a black dot in each panel, has a mass of $m_{\textrm{b}} = 0.1/2/8$ M$_{\textrm{Jup}}$ and a semi-major axis of $a_{\textrm{b}} = 8/10/20$ au. Planet c has a mass range of 0.001--20 M$_{\textrm{Jup}}$ and a semi-major axis range of 1--300 au. Hatched filled region are for dynamically unstable configurations using the criterion given in Equation~(\ref{eqn:stability}). The black dashed box in the top right corner represents roughly the parameter space GPI and SPHERE are sensitive to \citep[for a 30 Myr-old star at 30 pc;][]{bowl16}. The grey dashed box represents the predicted performance of E-ELT METIS L-band \citep[for a 30 Myr-old star at 30 pc;][]{quan15}. In most parameter space, planet c barely affects the forced eccentricity of the debris ring near to planet b.} \label{fig:ecc_ring}
\end{figure*}

\subsubsection{Inclination estimation error using a single-planet model}
In the last part of our exploration of the warp feature, we investigate how our interpretation of the detected planet's inclination is affected by the presence of an additional planet. We mentioned earlier that the disk inclination could be different from the single planet case if planet c were to have a different inclination from planet b. To study the strength of this effect, we build a two-planet model in which planet b has an inclination of 10$\degree$ ($i_{\textrm{b}} = 10\degree$), whereas planet c has an inclination of zero ($i_{\textrm{c}} = 0\degree$). We calculate the median forced inclinations of 2,000 evenly distributed planetesimals from 20--200 au and use it as a disk inclination indicator. The following equation is used to compare forced inclinations of the disk:
\begin{equation} \label{eqn:dev}
    \frac{\lvert i_{\textrm{disk}}' - i_{\textrm{disk}} \rvert}{i_{\textrm{disk}}} 100\%,
\end{equation}
where $i_{\textrm{disk}}'$ is the median disk forced inclination in systems with planet b and planet c and $i_{\textrm{disk}}$ is in systems with planet b only with $i_b = 10\degree$. As shown in Figure~\ref{fig:inc_err}, the disk forced inclination is reduced significantly when planet c is more massive than planet b. In other words, the median forced inclination of the disk is dominated by the inclination of the more massive planet. To illustrate this feature, we plot the disk forced inclination for configurations with different planet mass ratios in Figure~\ref{fig:inc_layer}. As planet c's mass increases, the disk forced inclination decreases from 10$\degree$ to 0$\degree$ (dark blue curve). It is noticeable that the singularity in the forced inclination shifts to greater semi-major axes as planet c's mass increases. This shift is mostly caused by a higher nodal precession rate of nearby particles as  a more massive planet c shifts the secular resonance to a greater distance. The triangle shaped feature in the contour in Figure~\ref{fig:inc_err} when planet c has a semi-major axis ($a_{\textrm{c}} > 70$ au) and low mass is caused by our definition on the outer edge of a debris disk as 200 au. Figure~\ref{fig:inc_tri} illustrates that as planet c's semi-major axis increases, the secular resonance occurs at a larger semi-major axis and its width increases. Because we truncate the disk at 200 au (i.e., shown as the vertical grey dash line in Figure~\ref{fig:inc_tri}), our calculation of the median forced inclination is affected by the high forced inclinations of planetesimals near to the resonance. In addition, as planet c's semi-major axis increases, the semi-major range of planetesimals with zero forced inclination increases. If we extend the disk outer edge to 500 au, we find the forced inclination of the planetesimals drops back to 10$\degree$, which still follows the inclination of the more massive planet. In interpreting observations, we should make use of the observed disk outer edge, on which the  disk median inclination depends.

In summary, for the warp feature, our inference on planet b's mass could be significantly flawed using a single-planet model if an undetected planet c is either massive or close to the warp. Disk morphology dissimilarities may let us distinguish between one-planet and two-planet models. The overall warped disk inclination follows the inclination of the more massive planet. The inference of the detected planet's inclination could be significantly different from the true value if a hidden planet is more massive than the detected planet. We also find secular resonances could affect the warp location and height and require detailed modeling for specific systems.

\subsection{Offset} \label{sec:offset}
In some exoplanetary systems \citep[e.g., Fomalhaut, HR 4796A;][]{kala05, macg17, tele00}, the geometric center of the debris disk or the debris ring has been observed to be offset from its central star. \citet{wyat99} explained the observed offset as a result of gravitational sculpting of a nearby planet on an eccentric orbit. The forced eccentricity ($e_{\text{forced}}$), a time-averaged eccentricity of a planetesimal at the semi-major axis $a$ from planet b, can be expressed as
\begin{equation}\label{eqn:ecc}
    e_{\text{forced}} = \frac{b_{3/2}^{(2)}(\alpha)}{b_{3/2}^{(1)}(\alpha)}e_{\text{b}},
\end{equation}
where $e_{\text{b}}$ is the posited planet b's eccentricity that generates the offset, $\alpha$ is the planet-planetesimal semi-major axis ratio in Equation (\ref{eqn: alpha_j}), and $b_{3/2}^{(1)}$ and $b_{3/2}^{(2)}$ are the Laplace coefficients in Equation~(\ref{eqn:coeff}). Unlike the forced inclination, the forced eccentricity depends on the semi-major axis ratio of the planetesimal to the planet (i.e., the forced eccentricity gets smaller when the planetesimal is more widely separated from the planet). Although the forced eccentricity itself is independent of the sculpting planet's mass,\footnote{The secular precession timescale still depends on the mass of planet b.} the semi-major axis and eccentricity of the posited planet can be inferred from the offset feature using the forced eccentricity equation. 

For the offset feature, we focus on the question: How does a second planet with an eccentricity different from the dominating planet affect the observed offset of the debris ring and the debris disk? To study planet c's effect on a debris ring, we consider two-planet configurations for which planet b has $e_{\textrm{b}} = 0.2$ and planet c has $e_{\textrm{c}} = 0$. We study the forced eccentricity of a ring near planet b and see how planet c's low eccentricity reduces the offset of that ring. In Figure~\ref{fig:ecc_ring}, we present the forced eccentricity reduction for an offset ring 2.5 Hill radii away from planet b caused by planet c. We place the debris ring 2.5 Hill radii outside planet b's orbit, assuming it is at the boundary of planet b's chaotic zone \citep{wisd80, dunc89, murr97}.\footnote{$\Delta a_{\textrm{chaotic}}\sim$ 1.3$\mu^{2/7} \approx 2.5R_{\textrm{Hill}}$ for Jupiter-mass planets.}
As shown in Figure~\ref{fig:ecc_ring}, in most parameter space, planet c barely affects the forced eccentricity of planetesimals near to planet b. Planet c only significantly affects the offset ring when planet b has a low mass (i.e., left panel) and planet c is both massive and close to the ring. The forced eccentricity of the ring in these configurations is reduced by at most 35\%. In other words, eccentricities of planetesimals are primarily dominated by the planet nearby.

We also generalize our study from a debris ring to a full debris disk. We investigate two planets with different eccentricities and their combined effect on the median forced eccentricity of two thousand planetesimals evenly distributed from 20 to 200 au. Because we study the forced eccentricity of the whole disk instead of a ring close to a planet, secular resonances become important and complicate the characterization of eccentricities. \citet{yelv18} recently investigated how planets' masses, semi-major axes, and eccentricities determine the locations, timescales, and widths of two exterior secular resonances in two-planet systems. In our study, two interior resonances and two exterior resonances all affect disk eccentricity. We find the median forced eccentricity of the disk can be strongly affected by secular resonances, making it challenging to characterize the planet eccentricity from the disk eccentricity. In general, the planet with greater semi-major axis dominates the disk eccentricity, assuming the system is old enough for the disk to have experienced one secular oscillation cycle.

In summary, for the offset feature, we find the forced eccentricity of a debris ring is usually primarily sculpted by the nearest planet. A disk's forced eccentricity is more complicated to model due to secular resonances, but the overall disk eccentricity follows the eccentricity of the planet with the greater semi-major axis inside or within the disk.

\section{Resonant Features in 2-planet Systems}\label{sec:resonant}
Debris disk features are sculpted by various types of planet-disk resonant interactions. A planet can remove nearby planetesimals, creating a gap, and sharpen the disk's edge because of the planet's overlapping resonances. In another type of resonant interaction, a planet captures planetesimals into its orbital resonances while migrating and redistributes the planetesimals to certain longitudes, creating an azimuthal asymmetry in the disk.  Properties of these resonant disk features (e.g., the gap width or the shape of clumpy features) can be used to constrain the sculpting planet's mass, semi-major axis, and migration history \citep{wisd80, quil06, wyat18}. Here we investigate if/how our inferences of these planet properties are compromised using a single-planet model when multiple planets reside in the system for edges and gaps (Section \ref{sec:edge}) and azimuthal asymmetries (Section \ref{sec:asymmetry}).

\subsection{Edges \& Gaps} 
\label{sec:edge}
Sharp edges and gaps have been observed in many debris disk systems \citep[e.g., Fomalhaut, HR 8799, HD 107146;][]{kala05,wiln18,ricc15}. The width of a gap and the sharpness of an edge can constrain the underlying planet's mass. \citet{wisd80} derived a simple relation of the width of the chaotic zone (i.e., a region where planetesimals are removed because of the planet's overlapping resonances) and the posited planet's mass for a planet on a circular orbit:
\begin{equation}\label{eqn:chaotic}
    \Delta a_{\textrm{chaotic}}/a_{\text{b}} \propto \mu^{2/7},
\end{equation}
where $\Delta a_{\textrm{chaotic}}$ is the width of the chaotic zone, $a_{\text{b}}$ is the sculpting planet's semi-major axis, and $\mu$ is the planet-star mass ratio. This relation is also known as the ``2/7" law. Follow-up studies took into account the planet's eccentricity \citep[e.g.,][]{glad93,chia09} and the planetesimal removal timescale \citep[e.g.,][]{morr15}. One of the most well-studied disks featuring edges and gaps is Fomalhaut's debris disk. \citet{quil06} applied the ``2/7" law to Fomalhaut and constrained the mass of Fomalhaut b (Fom b), the posited planet that would clear the chaotic zone and sharpen the inner edge of the belt. \citet{chia09} later constrained Fom b's mass, semi-major axis, and eccentricity using orbital constraints from the directly imaged planet \citep{kala08}. Unlike it in the $\beta$-Pic system, later studies showed the detection by \citet{kala08} might not be the planet responsible for the feature. On the one hand, \citet{chia09} still provides a good example of how to characterize the sculpting planet's mass from the width of the chaotic zone and the sharpness of the inner edge; on the other hand, Fom b is a cautionary tale about the dangers of assuming we know what planet is causing the features.

\subsubsection{Mass estimation error using a single-planet model}
\begin{figure}
\epsscale{1}
\plotone{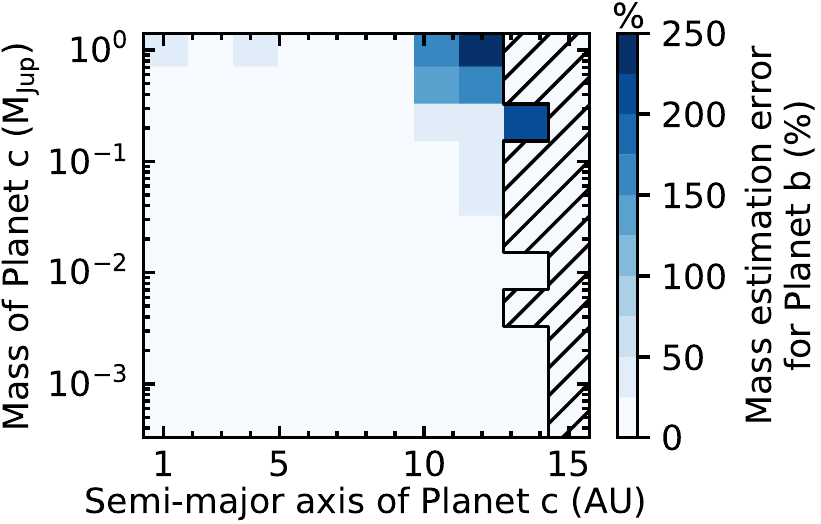}
\caption{Mass estimation error for planet b ($m_{\textrm{b}}$ = 1 M$_{\textrm{Jup}}$ and $a_{\textrm{b}}$ = 16 au) calculated from Equation~(\ref{eqn:error}) according to the width of the chaotic zone sculpted by both planet b and planet c. The hatched filled region represents dynamically unstable two-planet configurations over 10 Myr (i.e., e$_{\textrm{c}} >$ 1, collisions or ejections). The colorscale demonstrates levels of mass estimation error labeled by the colorbar. A planet c with an orbit close to the inner disk edge and with a comparable mass to planet b leads to the greatest mass estimation error for planet b.} \label{fig:edge_contour}
\end{figure}

If an edge or gap is sculpted by multiple planets, we expect the width of the chaotic zone to expand, and the sharpness of the inner edge may also be affected. In this section, we aim to answer the question: How would an undetected planet c impact our inference of planet b's mass from an edge or gap? To do so, we build a two-planet model including a Jupiter-mass planet b with a semi-major axis of 16 au ($m_{\textrm{b}}$ = 1 M$_{\textrm{Jup}}$ and $a_{\textrm{b}}$ = 16 au) and a planet c with a mass range of 0.0005--1 M$_{\textrm{Jup}}$ and a semi-major axis range of 1--15 au. This parameter space avoids planet c's chaotic zone exceeding planet b's chaotic zone and thus becoming the dominant planet. For simplicity, we also assume planet b and planet c have zero eccentricity and are coplanar to the disk. Resonant features, unlike secular features, do not have a known analytical solution in the four-body case (i.e., a host star, planet b, planet c, and a planetesimal). To investigate these features, we perform $N$-body simulations using \texttt{REBOUND} \citep{rein12,rein15a,rein15b} and \texttt{REBOUNDx} \citep{tama19}. For each two-planet configuration, we include two thousand test particles evenly distributed from 10--28 au with zero eccentricity and inclination and random mean anomalies. Each system is integrated for 5 Myr with a time step of one-percent of planet c's orbital period using the \texttt{WHFast} integrator \citep{rein15b}. The width of the chaotic zone is computed from the difference in semi-major axis between planet b and the median of the ten innermost planetesimals beyond planet b's orbit. We choose this metric because of its insensitivity to transient unstable test particles. We first apply the metric in the single-planet case and characterize how well it works. We consider a planet mass range of 1--5 M$_{\textrm{Jup}}$, characterize the widths of the chaotic zones after 5 Myr evolution, and fit the widths with Equation~(\ref{eqn:chaotic}). We find that the planet's mass scales with the the width of the chaotic zone as
\begin{equation}\label{eqn:width_single}
    \Delta a_{\textrm{chaotic}} = 1.75 \Big(\frac{m_{\textrm{b}}}{m_*}\Big)^{2/7} a_{\text{b}},
\end{equation}
where the coefficient 1.75 is from our fitting. The relation has some scatter and this variation can cause a small mass estimation error for the planet (i.e., $\lesssim$ 25\%). For two-planet configurations, the mass estimation error is calculated as the following:
\begin{equation} \label{eqn:error}
        \frac{\lvert m_{\textrm{b}}'-m_{\textrm{b}}\rvert}{m_{\textrm{b}}}100\%,
\end{equation}
where $m_{\textrm{b}}'$ is the estimated mass for planet b using Equation~(\ref{eqn:width_single}) and $m_{\textrm{b}}$ is planet b's true mass (m$_{\textrm{b}}$ = 1 M$_{\textrm{Jup}}$). In Figure~\ref{fig:edge_contour}, we present the mass estimation error for planet b in such two-planet configurations. The hatched filled region represents dynamically unstable two-planet configurations over 10 Myr (i.e., if e$_{\textrm{c}}>$ 1 or any ejections or collisions occur). As shown in Figure~\ref{fig:edge_contour}, planet c causes negligible mass estimation errors in most configurations due to its modest expansion of the chaotic zone. Only a planet c with a large semi-major axis and comparable mass to planet b leads to a relatively large estimation error (i.e., 100--250\%). In other words, our inference of planet b's mass may be flawed when an undetected planet c has a comparable mass and orbit to planet b.

\subsubsection{Further planet mass constraints from the edge sharpness}
The sharpness of the disk edge may place a further constraint on the planet mass. \citet{quil06} has shown that the eccentricity dispersion of planetesimals at the boundary of the chaotic zone is correlated with the sculpting planet's mass as $u_e \sim \mu^{3/7}$. For example, a low mass planet disperses planetesimals' eccentricities by a relatively small amount and thus sculpts a sharp edge, and vice versa for a massive planet. Given the observed sharpness of the disk edge, we may constrain the sculpting planet's mass. In a single-planet system, we find the observed edge sharpness is affected by planet b's mass via its effect on edge location and its perturbation of planetesimal eccentricities, in agree with \citet{quil06}. \citet{chia09} demonstrated the semi-major axis distribution of planetesimals could also affect the observed edge sharpness, because planetesimals on elliptical orbits at greater semi-major axes could cross the inner edge boundary and smooth the disk edge. To constrain the sculpting planet's mass from the observed sharpness of the disk edge, we have to include the contribution from planetesimals at greater semi-major axes in the model.

We conduct a series of case studies for two-planet configurations, aiming to explore an additional planet c's contribution to the edge sharpness. If planet c modifies the edge sharpness from that sculpted by planet b only, we can infer the need to use a multi-planet model instead of a single-planet one. For each selected configuration, we include four-thousand test particles near to the disk inner edge (i.e., 18--23 au) and integrate the system for 5 Myr using the \texttt{IAS15} integrator \citep[which treats close encounters more accurately than other integrators, e.g., \texttt{WHFast};][]{rein15a}. For two-planet configurations, we set $m_{\textrm{b}}$ = 1 M$_{\textrm{Jup}}$, $a_{\textrm{b}}$ = 16 au, $m_{\textrm{c}}$ = 0.05/0.5/1 M$_{\textrm{Jup}}$, and $a_{\textrm{c}}$ = 12 au. Planet c is placed close to planet b to have an observable effect on the disk edge. Kernel density estimation is applied to the semi-major axis and radial distance distribution of planetesimals. A Gaussian kernel with a bandwidth of 0.25 au is used.
\begin{figure}
\epsscale{1.}
\plotone{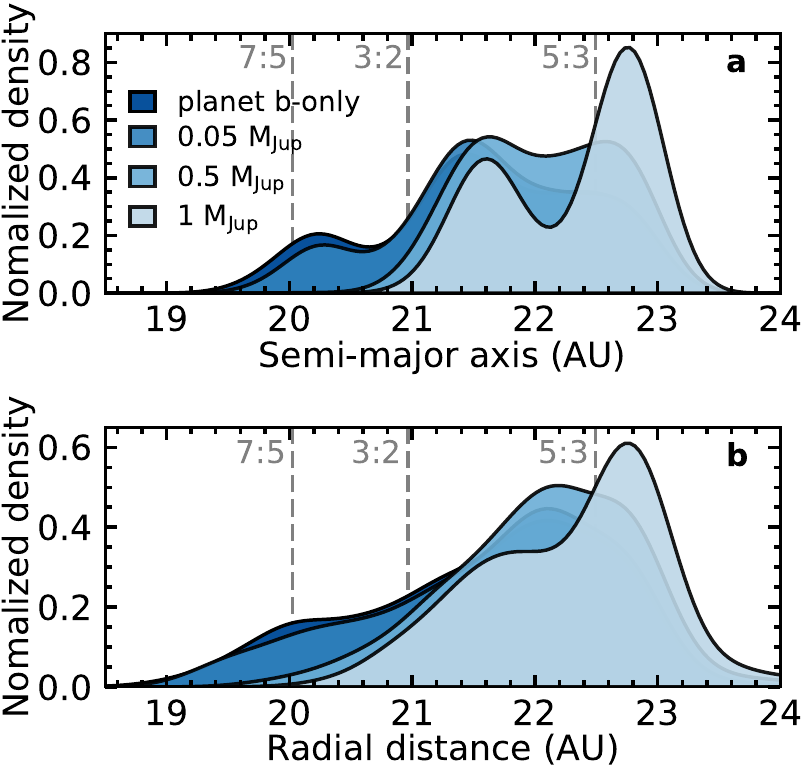}
\caption{Kernel density estimations for the planetesimals' semi-major axis distribution (upper panel) and stellar radial distance distribution (lower panel) sculpted by planet b-only (in darkest blue) or by both planet b and planet c with different masses for planet c (in lighter blues). We use Gaussian kernels with a bandwidth of 0.25 au. Planet b is a Jupiter-mass planet with a semi-major axis of 16 au. Planet c has $a_{\textrm{c}}$ = 12 au and $m_{\textrm{c}}$ = 0.05/0.5/1 M$_{\textrm{Jup}}$. Grey dashed lines indicate the location of mean-motion resonances of planetesimals with planet b.} \label{fig:edge_sharpness}
\end{figure}
As shown in the upper panel of Figure~\ref{fig:edge_sharpness}, increasing planet c's mass moves the disk edge to a greater semi-major axis. Each peak of semi-major axis distribution is slightly offset from locations of planet b's mean-motion resonances (i.e., labeled in grey dashed lines), because of planets' perturbation on the eccentricity of planetesimals (the higher the planet mass, the stronger the perturbation). We observe a slightly steeper inner edge slope for the planetesimal semi-major axis distribution when planet c is 0.5 or 1 Jupiter-mass, compared to planet c with 0.05 Jupiter-mass. This slightly sharpness change could be a result of a 0.5 or 1 Jupiter-mass planet c expanding the chaotic zone so the disk edge is close to planet b's 3:2 resonance; therefore this subtle steepness change is caused by the change in edge location rather than the eccentricity dispersion. However, in the radial distance distribution presented in the lower panel of Figure~\ref{fig:edge_sharpness}, the steepness change is less observable. The flatter slope in the radial distance distribution is because, as described by \citet{chia09}, planetesimals on elliptical orbits from greater semi-major axes cross the inner boundary and smooth the disk edge. It is difficult to distinguish among various two-planet configurations using the edge sharpness in the radial distribution of planetesimals, which is closer to what we observe than the semi-major axis distribution.

\begin{figure}
\epsscale{1.}
\plotone{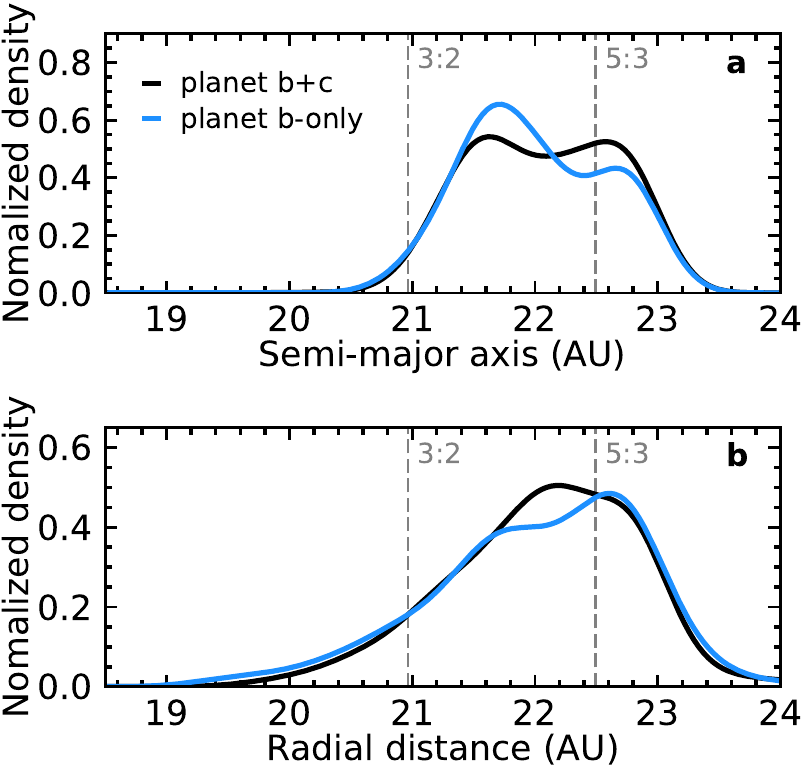}
\caption{Kernel density estimations for the planetesimals' semi-major axis distribution (upper panel) and stellar radial distance distribution (lower panel) sculpted by both planet b and planet c ($m_{\textrm{b}}$ = 1 M$_{\textrm{Jup}}$, $a_{\textrm{b}}$ = 16 au, $m_{\textrm{c}}$ = 0.5 M$_{\textrm{Jup}}$, $a_{\textrm{c}}$ = 12 au; in black) or by planet b-only ($m_{\textrm{b}}$ = 2 M$_{\textrm{Jup}}$, $a_{\textrm{b}}$ = 16 au; in blue). We use Gaussian kernels with a bandwidth of 0.25 au. Grey dashed lines indicate the location of mean-motion resonances of planetesimals with planet b. Edge sharpness dissimilarity is ambiguous in distinguishing between one-planet and two-planet systems.} \label{fig:sharpness_comp}
\end{figure}

We also compare the edge sharpness for two-planet configurations to single-planet configurations. We show one example in Figure~\ref{fig:sharpness_comp}. In the single-planet system, planet b is a 2 Jupiter-mass planet with a semi-major axis of 16 au ($m_{\textrm{b}}$ = 2 M$_{\textrm{Jup}}$, $a_{\textrm{b}}$ = 16 au). In the two-planet system, planet b is a Jupiter-mass planet with the same semi-major axis and planet c is a 0.5 Jupiter-mass planet with a semi-major axis of 12 au ($m_{\textrm{b}}$ = 1 M$_{\textrm{Jup}}$, $a_{\textrm{b}}$ = 16 au, $m_{\textrm{c}}$ = 0.5 M$_{\textrm{Jup}}$, $a_{\textrm{c}}$ = 12 au). The two configurations produce slightly different chaotic zone widths. However, as shown in Figure~\ref{fig:sharpness_comp}, we observed no significant difference in edge sharpness in either the planetesimal semi-major axis distribution or stellar radial distance distribution. We conclude edge sharpness dissimilarity cannot unambiguously distinguish between one-planet and two-planet systems.

If a planet has its mass estimated from the width of the chaotic zone, we could in principle model the disk edge using a single-planet model and compare the modeled edge sharpness to the observed one to validate the model. For example, imagine we directly imaged a planet with a semi-major axis of 16 au in a gap disk. Using the mass-chaotic zone width relation in Equation~(\ref{eqn:width_single}), we would infer the planet to be 3 Jupiter masses. If the planet is truly a Jupiter-mass planet with an interior companion, we would have overestimated that planet's mass by 200\%. However, a 3 Jupiter-mass planet would produce a greater planetesimal eccentricity dispersion compared to a Jupiter-mass planet with an interior companions, which would affect the edge sharpness. Nevertheless, we would still face the problem (i.e., \citealt{chia09}) that the semi-major axis distribution of planetesimals also affects the observed edge sharpness.

\subsubsection{The effects of two-planet in mean-motion resonance}
\begin{figure}
\epsscale{1.}
\plotone{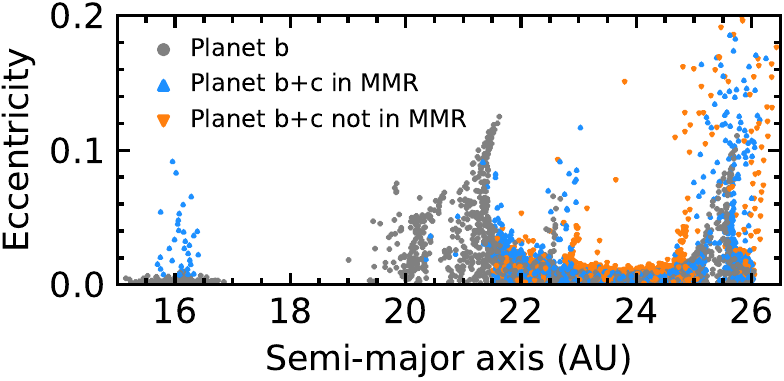}
\caption{Eccentricity versus semi-major axis of planetesimals in a system with planet b-only (grey dots), planet b and c in 2:1 mean motion resonance (blue triangle up), and planet b and c with a similar configuration but not in 2:1 mean motion resonance (orange triangle down). Planet b is a Jupiter-mass planet with a semi-major axis of 16 au in all three systems. Planet c is a Jupiter-mass planet with a semi-major axis of 10 au in both systems, having its resonant argument librating in the blue system but not in the orange system.} \label{fig:edge_mmr}
\end{figure}

We also investigate planet b and c's joint effect on the disk when they are in mean-motion orbital resonance (MMR). When planet c in resonance with planet b and at the same time planet b in resonance with outer planetesimals, a resonant chain is formed, which could affect planetesimals' orbits and stability. In Figure~\ref{fig:edge_mmr}, grey dots show the eccentricity versus semi-major axis of planetesimals in a planet b-only system ($m_{\textrm{b}}$ = 1 M$_{\textrm{Jup}}$ and $a_{\textrm{b}}$ = 16 au). Adding in a Jupiter-mass planet c with a semi-major axis of 10 au ($m_{\textrm{c}}$ = 1 M$_{\textrm{Jup}}$ and $a_{\textrm{c}}$ = 10 au) expands the chaotic zone from 19.5 au to 21.5 au, illustrated by blue and orange triangles. In the two-planet-in-resonance configuration (in blue), planet b and c are trapped in 2:1 MMR with the libration of the resonant argument $\varphi$ written as:
\begin{equation}
    \varphi = 2\lambda_{\textrm{b}} - \lambda_{\textrm{c}} - \varpi_{\textrm{b,c}},
\end{equation}
where $\lambda_{\textrm{b,c}}$ are mean longitudes of planet b and c and $\varpi_{\textrm{b,c}}$ are longitudes of pericenter of planet b and c. For the not-in-resonance configuration (in orange), planet b and c have similar masses and semi-major axes but are not librating in resonance. We observe no significant difference in the width of the chaotic zone or the sharpness of the edge between two types of configurations. Planet b and planet c in MMR can help to stabilize planetesimals at certain semi-major axes, e.g., planet b's trojans, compared to planet b and planet c not-in-resonance case.

\subsubsection{Summary of edge and gap features}
In summary, we explored how our inference of the posited planet's mass from the width of the chaotic zone and the edge sharpness can be affected by a hidden planet. For all two-planet configurations, the posited planet remains the dominant planet by construction. We find the posited planet's mass estimation may be flawed only for a hidden planet with a comparable mass and orbit to planet b, given the mass estimation from the width of the chaotic zone. (An additional planet has little effect on the sharpness of the planetesimals' radial distribution but may subtly affect the semi-major axis distribution.) In observations, two planets with similar masses and semi-major axes are likely to be both detected. In the unfortunate case where only one planet is detected, the mass estimation error for the detected planet could be as high as 250\%. Furthermore, planet b and planet c in MMR do not enhance the chaotic  behavior, compared to planet b and c near MMR.

\subsection{Azimuthal Asymmetries}
\label{sec:asymmetry}
Clumpy structures observed in debris disks \citep[e.g., $\beta$-Pic;][]{matr17, matr19} may result from resonant interactions between planets and planetesimals \citep[e.g.,][]{wyat03}. As a planet migrates inward/outward, it captures and redistributes planetesimals inside/outside its orbit to certain longitudes, resulting in disk asymmetries. \citet{wyat03} showed the probability of planetesimals being captured by a planet depends on the planet's mass, semi-major axis, and migration rate. Follow-up studies took into account additional factors such as a planet's eccentricity \citep[e.g.,][]{rech08}. Consequently, if an azimuthally asymmetric distribution of planetesimals is observed, we can place constraints on the underlying planet's mass, semi-major axis, and migration history.

\begin{figure}
\epsscale{1.1}
\plotone{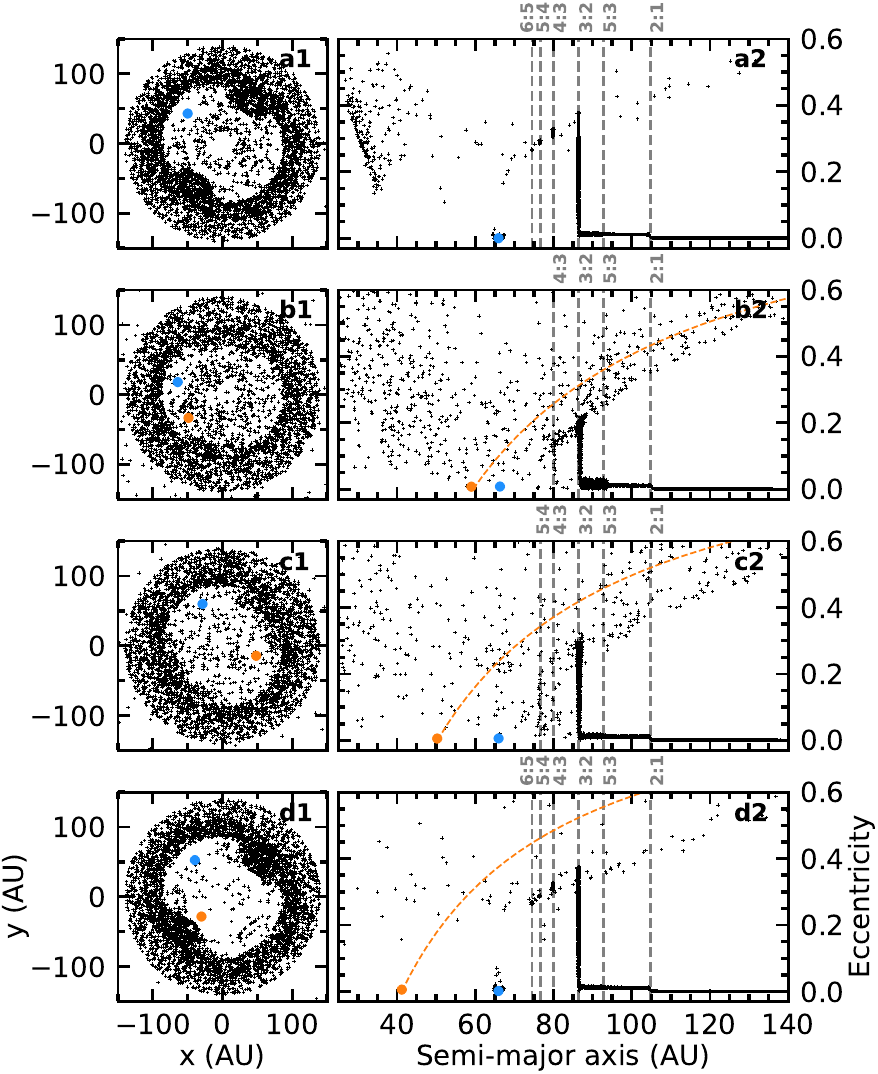}
\caption{Case studies of the azimuthal asymmetry feature for a single-planet configuration (panel a) and three two-planet configurations (panel b, c, and d). Left: planetesimal positions; right: eccentricity vs. semi-major axis distributions after 10 Myr of evolution. The blue and orange dots represent the final properties of planet b and planet c, respectively. Planets in all simulations are Neptune mass and migrate outward through the disk, following $a_0e^{t/20\textrm{Myr}}$. Grey dashed lines indicate locations of resonances with planet b. Orange dashed line demonstrates where periapses of planetesimals cross planet c's orbit (i.e., $e = 1-a_{\textrm{c}}/a$). Panel a presents a disk sculpted by the migration of planet b from 40 to 66 au. Panel b presents the same planet b along with the migration of planet c from 36 to 59.4 au; Panel c presents the same planet b along with the migration of planet c from 30.5 to 50.3 au; Panel d presents the same planet b along with the migration of planet c from 25 to 41.2 au.} \label{fig:clumps}
\end{figure}

A second planet, migrating along with the planet primarily sculpting the disk, may modify the resonant behavior of planetesimals and thus affect the asymmetries. To understand planet c's effect on the disk morphology, we investigate a few two-planet configurations including a Neptune-mass planet b and a Neptune-mass planet c. Both planets migrate outward through the disk by exchanging angular momentum with planetesimals \citep{fern84}. We consider a semi-major axis evolution of $a_0e^{t/20\textrm{Myr}}$ for both planets, where $a_0$ is the planet's initial semi-major axis and $t$ is time since migration began. With an initial semi-major axis of 40 au for planet b ($a_{0, \textrm{b}}$ = 40 au), this evolution means planet b migrates from 40 au to 66 au over 10 Myr. The migration rate we use here is motivated by \cite{wyat03} to optimize the capture probabilities into the 3:2 resonance for planetesimals. Three different initial semi-major axes of planet c are investigated: $a_{0, \textrm{c}}$ = 36 au, $a_{0, \textrm{c}}$ = 30.5 au, and $a_{0, \textrm{c}}$ = 25 au, which correspond to  final semi-major axes of 59.4 au, 50.3 au, and 41.2 au, respectively. We investigate how planet c with different semi-major axes affects the resonant behavior of planetesimals. We simulate planet migration using the \texttt{modify\_orbits\_forces} routine in the \texttt{REBOUNDx} library\footnote{\texttt{REBOUNDx(2.19.2)} \href{https://github.com/dtamayo/reboundx}{https://github.com/dtamayo/reboundx}} \citep{tama19}. We integrate 5000 planetesimals (i.e., massless test particles) evenly distributed from 40--140 au for 10 Myr and use the \texttt{IAS15} integrator to accurately handle close encounters. Our simulation outputs for the planet b-only configuration and three two-planet configurations are presented in Figure~\ref{fig:clumps}. 
Planetesimal positions (left panels) and eccentricity versus semi-major axis distributions of planetesimals (right panels) are shown. The blue and orange dots represent planet b and planet c, respectively. In the planet b-only configuration (a1, a2 in Figure~\ref{fig:clumps}), we find planetesimals outside planet b's orbits are captured in several resonances (e.g., 2:1, 5:3, 3:2, 4:3, 5:4, 6:5). A significant amount of planetesimals are captured into the 3:2 resonance with a wide range of eccentricities. Two symmetric clumps observed in Figure~\ref{fig:clumps} a1 are specifically these with high eccentricities (i.e., e $>$ 0.3). When planet c has a large semi-major axis (e.g., $a_{0, \textrm{c}}$ = 36 au or $a_{0, \textrm{c}}$ = 30.5 au; b1, b2 or c1, c2 in Figure~\ref{fig:clumps}), planetesimals captured in 6:5, 5:4, and 4:3 resonances are greatly reduced because the planetesimals cross the orbit of planet c. Although a large amount of planetesimals are still trapped in 3:2 resonance with planet b, highly eccentric planetesimals (e $>$ 0.3) are missing because the inner planet c destabilizes the outer eccentric planetesimals. Orange dashed lines indicate where periapses of planetesimals cross planet c's orbit (i.e., $e = 1-a_{\textrm{c}}/a$). In contrast, when planet c has a small semi-major axis ($a_{0, \textrm{c}}$ = 25 au; c1, c2 in Figure~\ref{fig:clumps}), planet c's orbit is too small to affect planetesimals trapped in resonance with planet b, even those on highly eccentric orbits. Meanwhile, planet c clears the inner region of the disk and makes the clumpy structures even easier to observe, compared to the planet b-only case. 

In summary, the azimuthal asymmetry feature is primarily dominated by a single planet. However, a second planet may affect the asymmetries by destabilizing planetesimals crossing its orbit. Disks with no observed clumpy structures could be the result of multiple migrating planets. 

We only explored a few case studies for a second planet with different semi-major axes. Although we placed two planets near to resonance for some configurations (e.g., Figure~\ref{fig:clumps} panel c near to 3:2 and panel d near to 2:1), in none of our case studies were the two planets captured into resonance with each other during migration, and this scenario's effect on the feature would be interesting to explore further. A full parameter space study of two-planet configurations (e.g., mass ratios, eccentricities, in MMR, etc.) is beyond the scope of this paper, but worthwhile to explore to fully understand a second planet's effect on azimuthal asymmetries.

\section{Synodic features in 2-planet Systems}
\label{sec:synodic}

A planetesimal's orbit can be modified by a nearby planet through close encounters. At each conjunction, the planetesimal's eccentricity and inclination are altered by the sculpting planet, along with a minor change in its semi-major axis. If a planet sculpts a ring of planetesimals, these planetesimals will show a distribution of eccentricities and inclinations, resulting in a radially thickened ring with a vertical scale height. We categorize this type of planet-disk interaction as a synodic feature. The underlying posited planet's mass and semi-major axis can be constrained from the normalized thickened ring width \citep[e.g.,][]{rodi14} and the vertical scale height \citep[e.g.,][]{quil07}. Both disk features are particularly useful for characterizing planet properties in systems with no detected planet. In this section, we investigate if/how using a single-planet model compromises our inferences of these planet properties when multiple planets reside in the system.

\subsection{Radially Thickened Rings} 
\label{sec:thickened}
As a planet clears nearby material to sculpt the disk edge (Section \ref{sec:edge}), it simultaneously stirs up planetesimals outside its chaotic zone; if the planet is near a ring of planetesimals (e.g., Fomalhaut or HR 4796A), we will observe a radially thickened ring. How much a ring gets thickened is directly related to the sculpting planet's mass and how close the planet is to the ring. With the assumption that the observed ring lies just beyond the sculpting planet's chaotic zone, we can infer the planet mass from the thickened ring width \citep{chia09, rodi14}. \citet{rodi14} confirmed a simple linear relation between the planet mass and the thickened ring width:
\begin{equation}
    \mu \propto \textrm{nFWHM},
\end{equation}
where $\mu$ is the planet-star mass ratio and nFWHM is the normalized full width at half maximum of the scattered light debris ring. Since we do not know the initial ring width for an observed system, this relationship places an upper limit on the planet's mass. 

We are interested in how a second planet in the system affects the ring width and consequently compromises our estimation of the first planet's mass. To begin with, we confirm the mass-normalized ring width relation given in \cite{rodi14} for a single-planet system. This step is to affirm that the linear relation found by \citet{rodi14} for scattered light also applies to planetesimal distributions. We consider a planet mass range of 0.3--10 M$_{\textrm{Jup}}$ and place a zero-width ring of two-thousand test particles outside the planet's orbit. Both the planet and planetesimals have zero eccentricity. The ring is placed as close to the planet as possible to have the strongest gravitational effect. The ring's semi-major axis is determined by the following criterion: we require more than 99\% of planetesimals must have a final semi-major axis between 0.1--10 of the initial semi-major axis but outside the sculpting planet's Hill sphere after 2000 orbit periods of the sculpting planet. This criterion, compared to \citet{rodi14}, is more stringent to make sure the debris ring is stable for at least Myr timescale (i.e., $\sim$50\% of planetesimals remains after 1 Myr). The ring width is computed as the range of the radial distances of the inner 99\% of planetesimals and further normalized by dividing by the semi-major axis of the ring center. We find the normalized ring width varies in time due to planet-planetesimal synodic and resonant interactions.\footnote{The variation follows the synodic timescale (i.e., the time taken for a conjunction), the repeated-conjunction timescale (i.e., the time taken to repeat conjunctions at the same location), and the resonance libration timescale.} To take into account the ring width variation, we take 80 evenly spaced snapshots of the system from 1000 to 1100 planet b's orbital period and use the median of the 80 normalized ring widths as the ring width for the modeling. We start at 1000 orbital period of planet b, adopted from \citet{rodi14}, and choose a time range of 100 planet b's orbital period since it is comparable to or longer than the synodic or resonance libration timescale for the planet mass range we study. \texttt{WHFast} integrator is used with a time step of one-percent of the planet's orbital period. For the planet mass range we study (0.3--10 M$_{\textrm{Jup}}$), a linear mass-normalized ring width relation is found:
\begin{equation}\label{eqn:width}
    w = 0.019\mu/\mu_{\textrm{Jup}}+0.082,
\end{equation}
where $w$ is the normalized ring width and $\mu/\mu_{\textrm{Jup}}$ is the planet-star mass ratio normalized by the Jupiter-Sun unit. The relation we find here is similar to what reported in \citet{rodi14} Equation (4). While a positive linear trend between the planet mass and the normalized ring width is evident, we find the linear relation has scatter, and the variations can cause a large mass estimation error for the planet (i.e., $\sim$150\%) even in the single planet case.

For two-planet configurations, we adapt a similar setup as the edge and gap feature (Section~\ref{sec:edge}): planet b is a Jupiter-mass planet with a semi-major axis of 16 au ($m_{\textrm{b}}$ = 1 M$_{\textrm{Jup}}$ and $a_{\textrm{b}}$ = 16 au); planet c is interior to planet b with a mass range of 0.0005--1 M$_{\textrm{Jup}}$ and a semi-major axis range of 1--15 au. We determine the stable outer ring locations and characterize the normalized ring widths following the procedures described for the single-planet case. The ring locations we find in these two-planet configurations are similar to the inner edge locations we find in Section~\ref{sec:edge}, as expected.

\begin{figure}
\epsscale{1.}
\plotone{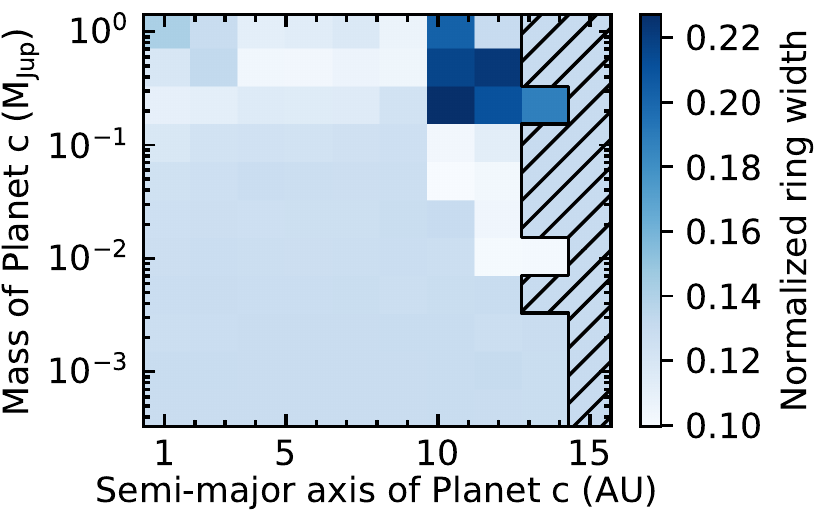}
\caption{Normalized ring width sculpted by both planet b ($m_{\textrm{b}}$ = 1 M$_{\textrm{Jup}}$ and $a_{\textrm{b}}$ = 16 au) and planet c. The hatched filled region represents dynamically unstable two-planet configurations over 10 Myr from $N$-body simulations (i.e., e$_{\textrm{c}} >$ 1) and is color-coded as the normalized ring width of 0.13, which is corresponding to the width sculpted by planet b only. The ring width is reduced in some two-planet configurations (i.e., colored in white/light blue) compared to a single-planet system and thickened for a planet c with a comparable mass and orbit to planet b.}  \label{fig:nFWHM}
\end{figure}

\begin{figure}
\epsscale{1.}
\plotone{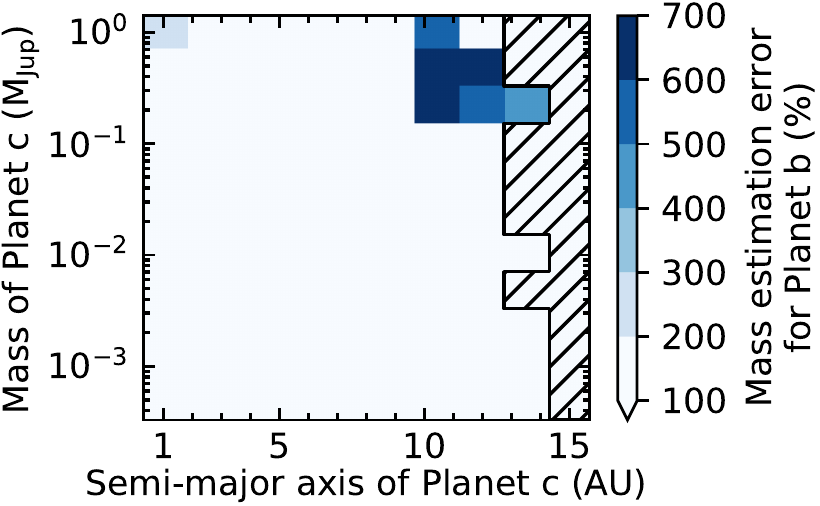}
\caption{Mass upper limit estimation error for planet b ($m_{\textrm{b}}$ = 1 M$_{\textrm{Jup}}$ and $a_{\textrm{b}}$ = 16 au) using the normalized ring width sculpted by both planet b and planet c. The hatched filled region represents dynamically unstable two-planet configurations over 10 Myr from $N$-body simulations (i.e., e$_{\textrm{c}} >$ 1). The colorscale demonstrates levels of mass estimation error labeled by the colorbar. Scatter in the single-planet relationship can cause a large mass estimation error for planet b ($\sim$ 150\%). The mass estimation error for planet b is severely flawed only for a planet c with a comparable mass and orbit to planet b.} \label{fig:ring_mass}
\end{figure}

We draw several conclusions about the normalized ring widths for two-planet configurations, which are summarized in Figure~\ref{fig:nFWHM}. First, if the planet c is a low mass planet with a small semi-major axis (i.e., the bottom-left corner), the normalized ring width is similar to the single-planet case. Secondly, the normalized ring width could be reduced by 8--23\% in systems with planet c with either a comparable mass or a comparable orbit to planet b but not both (i.e., configurations colored in white/light blue in Figure~\ref{fig:nFWHM}). However, for a Jupiter-mass planet c with a semi-major axis of 1 au, we find the ring width is further thickened compared to a single-planet model. The reason could be that the ring is placed closer to planet b for the configuration. Lastly, a planet c with a comparable mass and orbit to planet b (i.e., configurations colored in dark blue in Figure~\ref{fig:nFWHM}) significantly thickens the normalized ring width. However, we notice one exceptional two-planet configuration that causes less spread of the ring compared to other two-planet systems with similar configurations ($m_{\textrm{c}}$ = 1 M$_{\textrm{Jup}}$ and $a_{\textrm{c}}$ = 12 au; the upper right pixel in white). Future investigation is required to understand why the normalized ring width does not spread out as much in this configuration. The corresponding mass estimation errors for planet b in these two-planet configurations are presented in Figure~\ref{fig:ring_mass}. We estimate planet b's mass ($m_{\textrm{b}}'$) from Equation~(\ref{eqn:width}) and calculate the mass estimation error from Equation~(\ref{eqn:error}). Because of the large single-planet relationship intrinsic variation, we set the minimum mass estimation error percentage as 200\%.\footnote{The mass estimation error for a Jupiter-mass planet b with the normalized ring width of 0.13 is around 150\% using the mass-normalized ring width single-planet relation.} We reach a conclusion similar to that for edge and gap features: an overestimate of the mass upper limit for planet b occurs when a planet c with a comparable mass and orbit to planet b is present but undetected.

We also investigate how a planet c exterior to the ring affects the normalized ring width \citep[e.g., Fomalhaut's ring;][]{bole12}. We consider the same planet b ($m_{\textrm{b}}$ = 1 M$_{\textrm{Jup}}$ and $a_{\textrm{b}}$ = 16 au) and a Jupiter-mass or Neptune-mass planet c. We tune the semi-major axis of planet c and slightly increase the semi-major axis of the ring if necessary to attain stable system configurations. For a Jupiter-mass planet c, we find the normalized ring width could be thickened by $\sim$50\%, which corresponds to a mass estimation error of $\sim$500\% for planet b. In such configurations, planet c could even be tens of au apart from planet b but still have a strong gravitational effect on the ring width. For a Neptune-mass planet c, the normalized ring width could be either thickened or reduced by roughly 10--20\%, which may also induce a small mass estimation error for planet b. However, this error is comparable to the uncertainties of the single-planet relationship and may be considered negligible. 

In summary, the thickened ring width is usually primarily dominated by a single planet. However, the mass estimation for the posited planet (i.e., an upper limit on the planet's mass) could be severely flawed only for (1) an inner, hidden planet with a comparable mass and orbit to the posited planet or (2) an outer, wide-separation planet with a comparable mass to the posited planet. Moreover, the single-planet relationship intrinsic variations already causes a large mass estimation error for the posited planet and usually dominates over the error caused by an undetected planet.

\subsection{Scale Heights}
\label{sec:scale_height}
In Section \ref{sec:thickened}, we investigated the radially thickened ring feature, in which planets radially stir nearby planetesimals to a wider radial range. Similarly, embedded planets can also vertically stir nearby material, resulting in substantial scale heights \citep[e.g., AU Mic, $\beta$-Pic;][]{kris05, dale19, matr17}. Previous studies on scale height focused on how small stirring bodies (e.g., a Pluto-sized planetary embryo) imparts random kinetic energy to smaller planetesimals inside or near its Hill sphere \citep[e.g.,][]{quil07, theb07}. The scale height-to-radius aspect ratio (i.e., $H/r$) can be used as a mass or size indicator of the largest stirring bodies. In this work, we focus on scale heights sculpted by much larger stirring bodies -- planets. We study orbital changes of planetesimals (i.e., inclination changes) due to the close encounter with the sculpting planet and the resulting the disk scale height.

\citet{theb09} raised concerns about degeneracy in the scale height feature caused radiation pressure on small dust grains (i.e., with a size of 10 $\micro$m or less). Fortunately, this degeneracy only affects observations at short wavelengths (i.e., comparable to the small dust grain sizes). For longer wavelength debris disk observations (i.e., $\lambda$ $>$ 50 $\micro$m), the scale height will still be dominated by the stirring bodies.

We first investigate the scale height feature in a single-planet case. We are not aware of any analytical expressions that describe the scale height of planetesimals generated by a single, massive planet separated by several Hill radii from a ring. To understand how the planet's mass affects the ring's scale height, we use $N$-body simulations with a similar setup as the radially thickened ring feature. We consider a planet mass range of 0.3--10 M$_{\textrm{Jup}}$ and put a ring of test particles outside the sculpting planet's orbit with initial inclinations of 0.01 radians, random longitudes of the ascending nodes, and random mean anomalies. We study how a planet with zero inclination increases the disk aspect ratio from the initial value of 0.01. Although we find in general the scale height increases as the planet mass increases, there is not a monotonic relation between planet mass and scale height. The reason could be the planet's resonances affect the ring location, and the scale height is sensitive to the proximity of the planet. For example, for a 5 M$_{\textrm{Jup}}$ planet with a semi-major axis of 16 au, the scale height barely increases above the initial value of 0.01 because the ring needs to be 3.65 R$_{\textrm{Hill}}$ apart from the planet to avoid the planet's 5:3 resonance, which destabilizes the ring. The closest location for a stable ring is still too far from the planet to have its scale height effectively sculpted.
For a 6 M$_{\textrm{Jup}}$ planet, however, the scale height is greatly increased because a stable ring can be placed at a semi-major axis similar to the 5 M$_{\textrm{Jup}}$ planet (also avoiding the 5:3 resonance) but only 3.35 R$_{\textrm{Hill}}$ apart from the planet. 
Due to the planet's resonances and the sensitivity of the scale height to the proximity of the planet, we cannot derive a simple relation between the disk scale height and the planet mass. When interpreting observations, if both the planet and the ring location are known, we could characterize the sculpting planet mass from the width of the chaotic zone (e.g., Eqn~\ref{eqn:width_single}) and use that mass in numerical simulations to compare the simulated scale height to the observed one.

\begin{figure}
\epsscale{1.}
\plotone{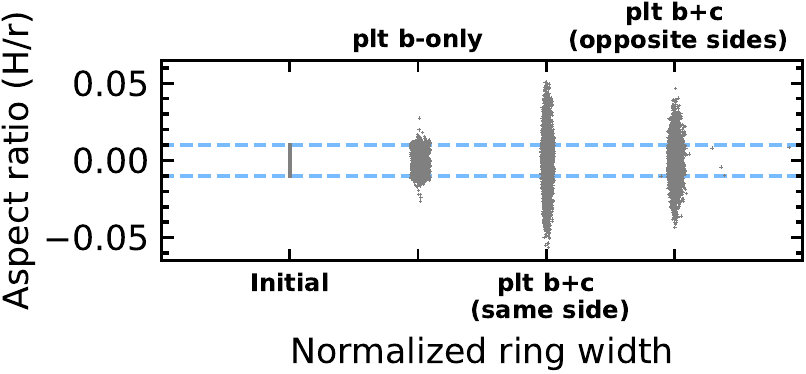}
\caption{Comparison of a debris ring's aspect ratio ($H/r$) and normalized ring width sculpted by planet b only (``plt b-only"; $m_{\textrm{b}}$ = 1 M$_{\textrm{Jup}}$, $a_{\textrm{b}}$ = 16 au), by planet b with a planet c interior to planet b's orbit (``plt b+c (same side)"; $m_{\textrm{c}}$ = 0.46 M$_{\textrm{Jup}}$, $a_{\textrm{c}}$ = 8.8 au), and by planet b with a planet c exterior to the debris ring (``plt b+c (opposite sides)"; $m_{\textrm{c}}$ = 0.054 M$_{\textrm{Jup}}$, $a_{\textrm{c}}$ = 32 au) after 1000 orbit periods of planet b. ``Initial" refers to the initial orbital parameters of the debris ring: 5000 test particles with a semi-major axis of 20 au and an inclination of 0.01 radians. Two blue dashed lines indicate an aspect ratio of $\pm$0.01 for reference. An additional planet can significantly increase the scale height, compared to the scale height sculpted by planet b only.}\label{fig:height}
\end{figure}

We further investigate how a second planet (i.e., planet c), in addition to the first planet (i.e., planet b), affects the scale height. If the scale height is increased by planet c, we are at risk of overestimating planet b's mass. In the single-planet configuration, planet b is a Jupiter-mass planet with a semi-major axis of 16 au ($m_{\textrm{b}}$ = 1 M$_{\textrm{Jup}}, a_{\textrm{b}}$ = 16 au), which is the same setup we used for the thickened ring feature (Section \ref{sec:thickened}) and the edge and gap feature (Section \ref{sec:edge}). We place a zero-width ring of five thousand test particles as close to planet b as possible to have the strongest gravitational effect on the ring while keep it stable ($a_{\textrm{ring}}$ = 20 au). We integrate the system for 1000 orbit periods of planet b with a time step of one-percent of planet b's orbital period using the \texttt{WHFast} integrator. In the left two panels of Figure~\ref{fig:height}, we plot the initial ring setup (i.e., a zero-width ring with an aspect ratio of 0.01) and the ring sculpted by the Jupiter-mass planet b only. In the planet b-only case, both the ring width and the scale height are moderately increased and the scale height is elevated from 0.01 to $\sim$0.015.

\begin{figure}
\epsscale{1.}
\plotone{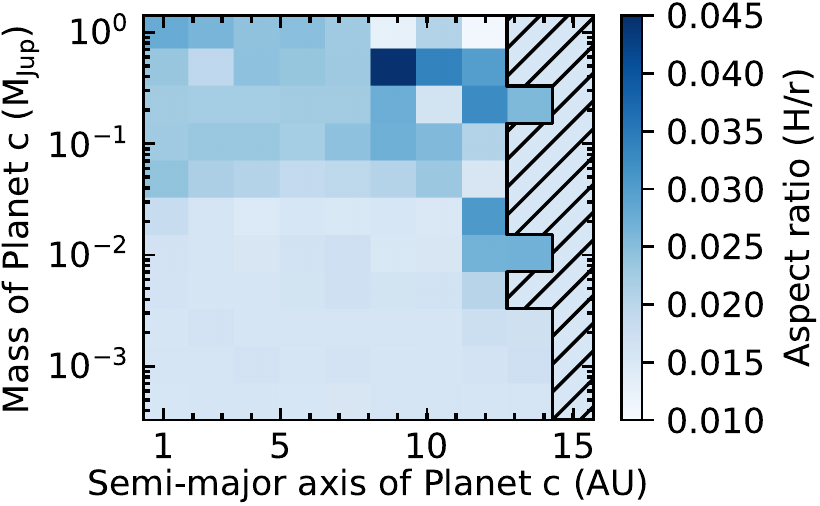}
\caption{Scale height-to-radius aspect ratio (H/r) sculpted by both planet b ($m_{\textrm{b}}$ = 1 M$_{\textrm{Jup}}$ and $a_{\textrm{b}}$ = 16 au) and planet c. The hatched filled region represents dynamically unstable two-planet configurations over 10 Myr from $N$-body simulations (i.e., e$_{\textrm{c}} >$ 1) and is color-coded relative to the aspect ratio of 0.015, which is corresponding to the scale height sculpted by planet b only. The aspect ratio could be moderately increased from 0.015 to roughly 0.02--0.025 for planet c more massive than 0.3 M$_{\textrm{Jup}}$.}  \label{fig:height_contour}
\end{figure}

We then study how a planet c in the same system as planet b further increases the scale height. Two types of two-planet configurations are investigated: (1) planet c is interior to planet b's orbit (i.e., two planets on the same side of the ring) and (2) planet c is exterior to the ring (i.e., two planets on the opposite sides of the ring). In both setups, we find configurations for which the scale height is significantly increase by adding in planet c, demonstrated in Figure~\ref{fig:height}. In the planet c-interior-to-planet b case, we explore planet c with a mass range of 0.0005--1 M$_{\textrm{Jup}}$ and a semi-major axis range of 1--15 au. Similar to the radially thickened ring feature, we take 80 evenly spaced snapshots of the system from 1000 to 1100 planet b's orbital period and use the median of the 80 aspect ratios as the aspect ratio for the modeling. As shown in Figure~\ref{fig:height_contour}, for planet c more massive than 0.03 Jupiter-mass ($m_{\textrm{c}} >$ 0.03 M$_{\textrm{Jup}}$ and $a_{\textrm{c}}$ = 1--15 au), the scale height could be moderately increased from 0.015 in the single planet case to roughly 0.02--0.025 in the two planet case. For a few configurations for which planet c is both massive and at a large semi-major axis, the scale height could be significantly increased to $\sim$0.05. 
We demonstrate one example in Figure~\ref{fig:height} as ``plt b+c (same side)", where $m_{\textrm{c}}$ = 0.46 M$_{\textrm{Jup}}$ and $a_{\textrm{c}}$ = 8.8 au. Planet b and the ring have the same setup as the single-planet configuration ($m_{\textrm{b}}$ = 1 M$_{\textrm{Jup}}$, $a_{\textrm{b}}$ = 16 au, and $a_{\textrm{ring}}$ = 20 au). For this configuration, planet c is massive and close to planet b but does not increase the width of the chaotic zone so that the ring is still close to planet b. The scale height is significantly increased to nearly 0.05 due to the gravitational perturbation from both planets. However, if the ring is sculpted by either planet b or planet c only, the scale height is barely enhanced. 

In the planet c-exterior-to-the-ring case, we consider a planet c of Jupiter-mass or Neptune-mass and a wide range of semi-major axes. In both cases, we find planet c could increase the scale height from that sculpted by planet b. Specifically, a Neptune-mass planet c close to the ring could significantly increase the scale height. We demonstrate this effect in Figure~\ref{fig:height} ``plt b+c (opposite sides)", where $m_{\textrm{c}}$ = 0.054 M$_{\textrm{Jup}}$ and $a_{\textrm{c}}$ = 32 au. Planet b and the ring have the same setup as the single-planet configuration ($m_{\textrm{b}}$ = 1 M$_{\textrm{Jup}}$, $a_{\textrm{b}}$ = 16 au, and $a_{\textrm{ring}}$ = 20 au). Compared to the scale height of $\sim$0.015 if sculpted by planet b only, the scale height is increase to $\sim$0.04, by adding a Neptune-mass planet c. For a Jupiter-mass planet c, the scale height may also be increased but usually only by a moderate amount because planet c could not be placed close to the ring without destabilizing the system.

For both types of two-planet configurations we explored, the proximity of the planets to the disk is key factor in increasing disk scale height. The sculpting planets' masses play a secondary role. The dependence on proximity explains why a Neptune-mass planet c has a stronger effect on the disk scale height than a Jupiter-mass planet c when two planets are on the opposite sides of the debris ring. The Neptune-mass planet can be closer to the ring without destabilizing it and therefore can increase the scale height more. Similarly, a planet c interior to planet b's orbit that does not increase the width of the chaotic zone has a stronger effect on the scale height, compared to those with higher masses and greater semi-major axes that will increase the width of the chaotic zone. 

In summary, the disk scale height can be used as a mass indicator of the stirring planet. Due to the complex interplay between Hill and resonant stability, there is no simple relationship between planet mass and disk scale height. Therefore deriving a mass requires detailed modeling of the individual system. Although the aspect ratio is usually primarily dominated by the nearby planet, we find a second stirring planet could affect the aspect ratio by a significant amount if it is close to the disk. For these configurations, our inference of the assumed planet's mass may be overestimated.

\section{Conclusion}\label{sec:conclusion}
Resolved debris disk features contain valuable information about the underlying planetary systems, and can reveal the mass, semi-major axis, and other orbital parameters of the planets by which they are created. Most existing planet-disk models assumed a single planet sculpts the disk feature. We investigated if/how planet properties inferred from single-planet models are compromised when a second, unseen planet resides in the system. On the bright side, we found debris disk features (warps, offsets, edges and gaps, azimuthal asymmetries, thickened rings, and scale heights) are usually primarily dominated by a single planet. However, the inferred planet properties can be severely flawed if the wrong planet is identified as the source of the morphology. We summarize our findings for each feature and highlight situations where we can infer the need to use a multi-planet model instead of a single-planet one from the disk morphology and thus avoid potential planet characterization errors.

\subsection{Warps}
The warp feature (Section~\ref{sec:warp}) allows for constraints on the sculpting planet's mass, semi-major axis, and inclination. If a planet is detected in the system and its semi-major axis is known, the planet's mass can be inferred from the warp location and the warp production time from Equation~(\ref{eqn:A}). We studied how our inference of the planet's mass could be compromised by an undetected planet present in the system. We found mass estimations for the detected planet could be significantly flawed when the detected planet is not the dominant planet in the system, demonstrated in Figure~\ref{fig:mass_error}. In extreme cases, the detected planet's mass could be misestimated by orders of magnitude. To avoid such estimation errors, we identified two types of disk morphology dissimilarities between a single-planet system and a two-planet system (Section~\ref{sec:dissimilarity}): the thickness of the outer disk (Figure~\ref{fig:warp_slope}) and multiple warps (Figure~\ref{fig:warps}). If the disk is well-resolved, comparing the single-planet model to the observed disk image allows one to infer whether or not a multi-planet model is needed.   

We also explored how an undetected planet could compromise our inference of the detected planet's inclination. We found a warped disk overall inclination follows the inclination of the more massive planet (Figure~\ref{fig:inc_err} and \ref{fig:inc_layer}). The inference of the detected planet’s inclination could be significantly different from the true value if a hidden planet is more massive than the detected planet. If an observed system has no detected planet, we can infer the inclination of the most massive planet from disk inclination. If the system has any detected planets, ruling out the presence of any more massive planets is necessary before characterizing detected planet's inclination from the disk. 

\subsection{Offsets}
The offset feature (Section~\ref{sec:offset}) is useful for inferring the sculpting planet's eccentricity. In the case where the observed ring lies just beyond the sculpting planet's chaotic zone, we can approximate the sculpting planet's eccentricity as the observed ring eccentricity (Equation~\ref{eqn:ecc}). We investigated two types of disk configurations: a debris ring and a debris disk. We found the forced eccentricity of a debris ring is primarily sculpted by the nearest planet, demonstrated in Figure~\ref{fig:ecc_ring}. If we can confirm the observed planet is the one sculpting the ring, we can infer the planet's eccentricity from the ring eccentricity. 

We found it is challenging to characterize the planet's eccentricity from the overall disk eccentricity. First, the disk eccentricity can be much smaller than the sculpting planet's eccentricity (Equation~\ref{eqn:ecc}). Secondly, the forced eccentricity of an entire disk in a multi-planet system can be complicated by secular resonances. A two-planet system would introduce two interior resonances and two exterior resonances all affect disk eccentricity. Fortunately, we may be able to identify a multi-planet system from disk features sculpted by secular resonances (e.g., multiple gaps; see \citealt{yelv18} for a detailed discussion).

\subsection{Edges \& Gaps}
Width and edge features (Section~\ref{sec:edge}) place constraints on the sculpting planet's mass and semi-major axis. We confirm the relation between planet mass and the width of the chaotic zone, also known as the ``2/7" law (Equation~\ref{eqn:width_single}). For two-planet configurations, we studied how a second, hidden planet  would further increase the width of the chaotic zone and thus impact our inference of the detected planet's mass. We found the chaotic zone width is usually primarily dominated by a single planet. The mass estimation for the detected planet is flawed only for a hidden planet with a comparable mass and orbit to the detected planet, as shown in Figure~\ref{fig:edge_contour}. The mass estimation error could at most be $\sim$250\% in such cases. Fortunately, in observed systems we would expect two planets with similar masses and semi-major axes be both detected (via either direct imaging or radial velocity) if one is detected. We can be less concerned about the planet's mass estimation if we can rule out another, more massive planet closer to the ring.   

The gap's edge sharpness is affected by the planet's mass \citep{quil06}, but it is difficult to constrain the planet's mass uniquely even in the single planet case \citep{chia09}. We further explored how a second planet modifies the edge sharpness and found edge sharpness dissimilarity cannot unambiguously distinguish between one-planet and two-planet systems, demonstrated in Figure~\ref{fig:edge_sharpness} and \ref{fig:sharpness_comp}.

We also investigated how would two planets in mean-motion resonance, compared to two similarly spaced, similar mass planets not in mean-motion resonance, affect the disk morphology. We found two planets in MMR do not significantly enhance the chaotic behavior of the planetesimals or modify the edge sharpness, demonstrated in Figure~\ref{fig:edge_mmr}.

\subsection{Azimuthal Asymmetries}
The azimuthal asymmetry feature (Section~\ref{sec:asymmetry}) places constraints on the underlying planet's mass, semi-major axis, and migration history. We studied how a second planet, migrating along with the planet primarily sculpting the disk, modifies the resonant behavior of planetesimals and thus affect the asymmetries. We found a second planet could affect the azimuthal asymmetries by destabilizing captured planetesimals crossing its orbit, as shown in Figure~\ref{fig:clumps} panel b and c. If no clumpy structure is observed in a system, we cannot rule out migration: the lack of resonant objects could be the result of multiple migrating planets. However, if clumpy structure is observed, it is likely to be sculpted by one dominant planet (e.g., Figure~\ref{fig:clumps} panel d). Using a single-planet model to charaterize that planet's properties and migration history is generally acceptable.

\subsection{Thickened Rings}
The thickened ring feature (Section \ref{sec:thickened}) places an upper limit on the sculpting planet’s mass, which is particularly useful for characterizing planet properties in systems with no detected planet. We confirmed the planet mass-normalized ring width relation derived in \citep{rodi14} for scattered light images also works for planetesimal distributions for single-planet systems, described in Equation (\ref{eqn:width}). We investigated two types of two-planet configurations: two planets on the same side of the ring and two planets on the opposite sides of the ring. We found the thickened ring width is usually primarily dominated by the more massive planet closer to the ring. For configurations with both planets interior to the ring, the ring width could be  thickened by 40--70\% if two planets have comparable mass and orbits (Figure~\ref{fig:nFWHM}). Using a single-planet model for such configurations, we will greatly overestimate the upper limit on the sculpting planet's mass by 400--650\% (Figure~\ref{fig:ring_mass}). For configurations with both planets on the opposite side of the ring, if two planets have comparable mass, the ring could also get significantly thickened even if one planet is far from the ring. Ignoring such hidden planets could cause a mass estimation error of $\lesssim$500\%. For both types of two-planet configurations, a second planet could further thicken the ring width and lead to an incorrect inference on the assumed planet's maximum mass. When interpreting observations, we must be cautious using the planet mass-normalized ring width relation when we do not know how many planets may contribute to the observed ring feature. However, if the ring system we study has one or more detected planets likely sculpting the ring, we can be less worried because another planet can only compromise our inference if it has comparable mass and comparable/greater semi-major axis. If one is detected via either direct imaging or radial velocity, the other one is likely to be detected too.

\subsection{Scale Heights}
The scale height feature (Section~\ref{sec:scale_height}) is useful to characterize the mass of the sculpting planet. We examined how a second planet would further increase the disk scale height and thus led to an overestimation for the assumed planet's mass. We explored two types of two-planet configurations: two planets on the same side of the debris ring (Figure~\ref{fig:height_contour}) and two planets on the opposite sides of the ring. We found the scale height is usually primarily dominated by the stirring planet with greater mass and closer to the ring. However, in both types of configurations, we found systems for which the scale height is significantly increased by the second planet, demonstrated in Figure~\ref{fig:height}. The proximity of the planets to the ring is the key factor that increases scale height.

\subsection{Summary}
Using our findings, we rank the risk levels of using a single-planet model to characterize planet properties for disk features we studied. We are at high risk of misinterpreting planet properties from disk warps; at moderate risk from disk edges and gaps, radially thickened rings, and scale height features; and at low risk from host star-disk center offsets and azimuthal asymmetries.

\acknowledgments
We thank the referee for a helpful report that improved the clarity of the paper. We are grateful to Meredith Hughes for helpful discussion and feedback. We thank Ruobing Dong, Tom Esposito, Brad Foley, Eric Ford, Kevin Luhman, and Jason Wright for helpful comments. The Center for Exoplanets and Habitable Worlds is supported by the Pennsylvania State University, the Eberly College of Science, and the Pennsylvania Space Grant Consortium. We gratefully acknowledge support from the the Alfred P. Sloan Foundation's Sloan Research Fellowship and the Oak Ridge Associated Universities Ralph E. Powe Junior Faculty Enhancement Award. AS is supported by funding from the European Research Council under the European Community’s H2020 (2014-2020/ERC Grant Agreement No. 669416 ‘LUCKY STAR’). Computations for this research were performed on the Pennsylvania State University’s Institute for CyberScience Advanced CyberInfrastructure (ICS-ACI). This content is solely the responsibility of the authors and does not necessarily represent the views of the Institute for CyberScience. Simulations in this paper made use of the REBOUND code which can be downloaded freely at \href{http://github.com/hannorein/rebound}{http://github.com/hannorein/rebound}.

\software{\texttt{REBOUND} \citep{rein12, rein15a,rein15b}, \texttt{REBOUNDx} \citep{tama19}, \texttt{Matplotlib} \citep{hunt07, droe16}, \texttt{Numpy} \citep{vand11}, \texttt{Jupyter} \citep{kluy16}}

\bibliographystyle{aasjournal}
\bibliography{ref}

\begin{thebibliography}{}
\expandafter\ifx\csname natexlab\endcsname\relax\def\natexlab#1{#1}\fi
\providecommand{\url}[1]{\href{#1}{#1}}

\bibitem[{{Augereau} {et~al.}(2001){Augereau}, {Nelson}, {Lagrange},
  {Papaloizou}, \& {Mouillet}}]{auge01}
{Augereau}, J.~C., {Nelson}, R.~P., {Lagrange}, A.~M., {Papaloizou}, J.~C.~B.,
  \& {Mouillet}, D. 2001, \aap, 370, 447

\bibitem[{{Bastien} {et~al.}(2014){Bastien}, {Stassun}, \& {Pepper}}]{bast14}
{Bastien}, F.~A., {Stassun}, K.~G., \& {Pepper}, J. 2014, \apj, 788, L9

\bibitem[{{Beichman} {et~al.}(2019){Beichman}, {Barrado}, {Belikov}, {Biller},
  {Boccaletti}, {Burrows}, {Danielski}, {Choquet}, {Doyon}, {Fortney},
  {Gaspar}, {Glasse}, {Hinkley}, {Hu}, {Kataria}, {Krist}, {Lafreni{\`e}re},
  {Lagage}, {Lunine}, {Marley}, {Mawet}, {Meshkat}, {Meyer}, {Oppenheimer},
  {Perrin}, {Pueyo}, {Ressler}, {Rieke}, {Rieke}, {Roellig}, {Serabyn},
  {Schlieder}, {Skemer}, {Soummer}, {Su}, {Tremblin}, {Venot}, \&
  {Ygouf}}]{beic19}
{Beichman}, C., {Barrado}, D., {Belikov}, R., {et~al.} 2019, \baas, 51, 58

\bibitem[{{Biller} {et~al.}(2015){Biller}, {Liu}, {Rice}, {Wahhaj}, {Nielsen},
  {Hayward}, {Kuchner}, {Close}, {Chun}, {Ftaclas}, \& {Toomey}}]{bill15}
{Biller}, B.~A., {Liu}, M.~C., {Rice}, K., {et~al.} 2015, \mnras, 450, 4446

\bibitem[{{Boley} {et~al.}(2012){Boley}, {Payne}, {Corder}, {Dent}, {Ford}, \&
  {Shabram}}]{bole12}
{Boley}, A.~C., {Payne}, M.~J., {Corder}, S., {et~al.} 2012, \apjl, 750, L21

\bibitem[{{Bonnefoy} {et~al.}(2014){Bonnefoy}, {Marleau}, {Galicher}, {Beust},
  {Lagrange}, {Baudino}, {Chauvin}, {Borgniet}, {Meunier}, {Rameau},
  {Boccaletti}, {Cumming}, {Helling}, {Homeier}, {Allard}, \&
  {Delorme}}]{bonn14}
{Bonnefoy}, M., {Marleau}, G.~D., {Galicher}, R., {et~al.} 2014, \aap, 567, L9

\bibitem[{{Borucki} {et~al.}(2010){Borucki}, {Koch}, {Basri}, {Batalha},
  {Brown}, {Caldwell}, {Caldwell}, {Christensen-Dalsgaard}, {Cochran}, \&
  {DeVore}}]{boru10}
{Borucki}, W.~J., {Koch}, D., {Basri}, G., {et~al.} 2010, Science, 327, 977

\bibitem[{{Bowler}(2016)}]{bowl16}
{Bowler}, B.~P. 2016, \pasp, 128, 102001

\bibitem[{{Brande} {et~al.}(2019){Brande}, {Barclay}, {Schlieder}, {Lopez}, \&
  {Quintana}}]{bran19}
{Brande}, J., {Barclay}, T., {Schlieder}, J.~E., {Lopez}, E.~D., \& {Quintana},
  E.~V. 2019, arXiv e-prints, arXiv:1911.02022

\bibitem[{{Brandl} {et~al.}(2014){Brandl}, {Feldt}, {Glasse}, {Guedel},
  {Heikamp}, {Kenworthy}, {Lenzen}, {Meyer}, {Molster}, {Paalvast}, {Pantin},
  {Quanz}, {Schmalzl}, {Stuik}, {Venema}, \& {Waelkens}}]{bran14}
{Brandl}, B.~R., {Feldt}, M., {Glasse}, A., {et~al.} 2014, Society of
  Photo-Optical Instrumentation Engineers (SPIE) Conference Series, Vol. 9147,
  {METIS: the mid-infrared E-ELT imager and spectrograph}, 914721

\bibitem[{{Brandl} {et~al.}(2018){Brandl}, {Absil}, {Ag{\'o}cs}, {Baccichet},
  {Bertram}, {Bettonvil}, {van Boekel}, {Burtscher}, {van Dishoeck}, {Feldt},
  {Garcia}, {Glasse}, {Glauser}, {G{\"u}del}, {Haupt}, {Kenworthy}, {Labadie},
  {Laun}, {Lesman}, {Pantin}, {Quanz}, {Snellen}, {Siebenmorgen}, \& {van
  Winckel}}]{bran18}
{Brandl}, B.~R., {Absil}, O., {Ag{\'o}cs}, T., {et~al.} 2018, in Society of
  Photo-Optical Instrumentation Engineers (SPIE) Conference Series, Vol. 10702,
  \procspie, 107021U

\bibitem[{{Chen} {et~al.}(2019){Chen}, {Ballering}, {Duchene}, {Gaspar},
  {Kolokolova}, {Lisse}, {Mazoyer}, {Moro-Martin}, {Ren}, {Su}, \&
  {Wyatt}}]{chen19}
{Chen}, C., {Ballering}, N., {Duchene}, G., {et~al.} 2019, \baas, 51, 342

\bibitem[{{Chiang} {et~al.}(2009){Chiang}, {Kite}, {Kalas}, {Graham}, \&
  {Clampin}}]{chia09}
{Chiang}, E., {Kite}, E., {Kalas}, P., {Graham}, J.~R., \& {Clampin}, M. 2009,
  \apj, 693, 734

\bibitem[{{Cumming} {et~al.}(2008){Cumming}, {Butler}, {Marcy}, {Vogt},
  {Wright}, \& {Fischer}}]{cumm08}
{Cumming}, A., {Butler}, R.~P., {Marcy}, G.~W., {et~al.} 2008, \pasp, 120, 531

\bibitem[{{Daley} {et~al.}(2019){Daley}, {Hughes}, {Carter}, {Flaherty},
  {Lambros}, {Pan}, {Schlichting}, {Chiang}, {Wyatt}, {Wilner}, {Andrews}, \&
  {Carpenter}}]{dale19}
{Daley}, C., {Hughes}, A.~M., {Carter}, E.~S., {et~al.} 2019, arXiv e-prints,
  arXiv:1904.00027

\bibitem[{{Dawson} \& {Johnson}(2018)}]{daws18}
{Dawson}, R.~I., \& {Johnson}, J.~A. 2018, Annual Review of Astronomy and
  Astrophysics, 56, 175

\bibitem[{{Dawson} {et~al.}(2011){Dawson}, {Murray-Clay}, \&
  {Fabrycky}}]{daws11}
{Dawson}, R.~I., {Murray-Clay}, R.~A., \& {Fabrycky}, D.~C. 2011, \apj, 743,
  L17

\bibitem[{{Debes} {et~al.}(2019){Debes}, {Choquet}, {Faramaz}, {Duchene},
  {Hines}, {Stark}, {Ygouf}, {Girard}, {Moro-Martin}, {Arriaga}, {Chen},
  {Currie}, {Dodson-Robinson}, {Douglas}, {Kalas}, {Lisse}, {Mawet}, {Mazoyer},
  {Mennesson}, {Millar-Blanchaer}, {Sivramakrishnan}, \& {Wang}}]{debe19}
{Debes}, J., {Choquet}, E., {Faramaz}, V.~C., {et~al.} 2019, \baas, 51, 566

\bibitem[{{Dent} {et~al.}(2014){Dent}, {Wyatt}, {Roberge}, {Augereau},
  {Casassus}, {Corder}, {Greaves}, {de Gregorio-Monsalvo}, {Hales}, {Jackson},
  {Hughes}, {Lagrange}, {Matthews}, \& {Wilner}}]{dent14}
{Dent}, W.~R.~F., {Wyatt}, M.~C., {Roberge}, A., {et~al.} 2014, Science, 343,
  1490

\bibitem[{{Droettboom} {et~al.}(2016){Droettboom}, {Hunter}, {Caswell},
  {Firing}, {Nielsen}, {Elson}, {Root}, {Dale}, {Lee}, {Sepp{\"a}nen},
  {McDougall}, {Straw}, {May}, {Varoquaux}, {Yu}, {Ma}, {Moad}, {Silvester},
  {Gohlke}, {W{\"u}rtz}, {Hisch}, {Ariza}, {Cimarron}, {Thomas}, {Evans},
  {Ivanov}, {Whitaker}, {Hobson}, {mdehoon}, \& {Giuca}}]{droe16}
{Droettboom}, M., {Hunter}, J., {Caswell}, T.~A., {et~al.} 2016, {Matplotlib:
  Matplotlib V1.5.1}, vv1.5.1,  Zenodo, doi:10.5281/zenodo.44579

\bibitem[{{Duncan} {et~al.}(1989){Duncan}, {Quinn}, \& {Tremaine}}]{dunc89}
{Duncan}, M., {Quinn}, T., \& {Tremaine}, S. 1989, \icarus, 82, 402

\bibitem[{{Fernandez} \& {Ip}(1984)}]{fern84}
{Fernandez}, J.~A., \& {Ip}, W.~H. 1984, \icarus, 58, 109

\bibitem[{{Gladman}(1993)}]{glad93}
{Gladman}, B. 1993, \icarus, 106, 247

\bibitem[{{Howard} {et~al.}(2010){Howard}, {Marcy}, {Johnson}, {Fischer},
  {Wright}, {Isaacson}, {Valenti}, {Anderson}, {Lin}, \& {Ida}}]{howa10}
{Howard}, A.~W., {Marcy}, G.~W., {Johnson}, J.~A., {et~al.} 2010, Science, 330,
  653

\bibitem[{{Hughes} {et~al.}(2018){Hughes}, {Duch{\^e}ne}, \&
  {Matthews}}]{hugh18}
{Hughes}, A.~M., {Duch{\^e}ne}, G., \& {Matthews}, B.~C. 2018, Annual Review of
  Astronomy and Astrophysics, 56, 541

\bibitem[{{Hunter}(2007)}]{hunt07}
{Hunter}, J.~D. 2007, Computing in Science and Engineering, 9, 90

\bibitem[{{Kalas} {et~al.}(2005){Kalas}, {Graham}, \& {Clampin}}]{kala05}
{Kalas}, P., {Graham}, J.~R., \& {Clampin}, M. 2005, \nat, 435, 1067

\bibitem[{{Kalas} {et~al.}(2008){Kalas}, {Graham}, {Chiang}, {Fitzgerald},
  {Clampin}, {Kite}, {Stapelfeldt}, {Marois}, \& {Krist}}]{kala08}
{Kalas}, P., {Graham}, J.~R., {Chiang}, E., {et~al.} 2008, Science, 322, 1345

\bibitem[{{Kluyver} {et~al.}(2016){Kluyver}, {Ragan-Kelley}, {Pérez},
  {Granger}, {Bussonnier}, \& {Frederic}}]{kluy16}
{Kluyver}, T., {Ragan-Kelley}, B., {Pérez}, F., {et~al.} 2016, {Jupyter
  Notebooks - a publishing format for reproducible computational workflows}, 87

\bibitem[{{Konishi} {et~al.}(2016){Konishi}, {Grady}, {Schneider}, {Shibai},
  {McElwain}, {Nesvold}, {Kuchner}, {Carson}, {Debes}, {Gaspar}, {Henning},
  {Hines}, {Hinz}, {Jang-Condell}, {Moro-Mart{\'\i}n}, {Perrin}, {Rodigas},
  {Serabyn}, {Silverstone}, {Stark}, {Tamura}, {Weinberger}, \&
  {Wisniewski}}]{koni16}
{Konishi}, M., {Grady}, C.~A., {Schneider}, G., {et~al.} 2016, \apj, 818, L23

\bibitem[{{Krist} {et~al.}(2005){Krist}, {Ardila}, {Golimowski}, {Clampin},
  {Ford}, {Illingworth}, {Hartig}, {Bartko}, {Ben{\'\i}tez}, {Blakeslee},
  {Bouwens}, {Bradley}, {Broadhurst}, {Brown}, {Burrows}, {Cheng}, {Cross},
  {Demarco}, {Feldman}, {Franx}, {Goto}, {Gronwall}, {Holden}, {Homeier},
  {Infante}, {Kimble}, {Lesser}, {Martel}, {Mei}, {Menanteau}, {Meurer},
  {Miley}, {Motta}, {Postman}, {Rosati}, {Sirianni}, {Sparks}, {Tran},
  {Tsvetanov}, {White}, \& {Zheng}}]{kris05}
{Krist}, J.~E., {Ardila}, D.~R., {Golimowski}, D.~A., {et~al.} 2005, \aj, 129,
  1008

\bibitem[{{Lagrange} {et~al.}(2012){Lagrange}, {De Bondt}, {Meunier},
  {Sterzik}, {Beust}, \& {Galland}}]{lagr12}
{Lagrange}, A.~M., {De Bondt}, K., {Meunier}, N., {et~al.} 2012, \aap, 542, A18

\bibitem[{{Lagrange} {et~al.}(2009){Lagrange}, {Gratadour}, {Chauvin}, {Fusco},
  {Ehrenreich}, {Mouillet}, {Rousset}, {Rouan}, {Allard}, {Gendron}, {Charton},
  {Mugnier}, {Rabou}, {Montri}, \& {Lacombe}}]{lagr09}
{Lagrange}, A.~M., {Gratadour}, D., {Chauvin}, G., {et~al.} 2009, \aap, 493,
  L21

\bibitem[{{Lagrange} {et~al.}(2010){Lagrange}, {Bonnefoy}, {Chauvin}, {Apai},
  {Ehrenreich}, {Boccaletti}, {Gratadour}, {Rouan}, {Mouillet}, {Lacour}, \&
  {Kasper}}]{lagr10}
{Lagrange}, A.~M., {Bonnefoy}, M., {Chauvin}, G., {et~al.} 2010, Science, 329,
  57

\bibitem[{{Lagrange} {et~al.}(2019){Lagrange}, {Meunier}, {Rubini}, {Keppler},
  {Galland}, {Chapellier}, {Michel}, {Balona}, {Beust}, {Guillot}, {Grandjean},
  {Borgniet}, {M{\'e}karnia}, {Wilson}, {Kiefer}, {Bonnefoy}, {Lillo-Box},
  {Pantoja}, {Jones}, {Iglesias}, {Rodet}, {Diaz}, {Zapata}, {Abe}, \&
  {Schmider}}]{lagr19}
{Lagrange}, A.~M., {Meunier}, N., {Rubini}, P., {et~al.} 2019, Nature
  Astronomy, 421

\bibitem[{{Lee} \& {Chiang}(2016)}]{lee16}
{Lee}, E.~J., \& {Chiang}, E. 2016, \apj, 827, 125

\bibitem[{{MacGregor} {et~al.}(2013){MacGregor}, {Wilner}, {Rosenfeld},
  {Andrews}, {Matthews}, {Hughes}, {Booth}, {Chiang}, {Graham}, {Kalas},
  {Kennedy}, \& {Sibthorpe}}]{macg13}
{MacGregor}, M.~A., {Wilner}, D.~J., {Rosenfeld}, K.~A., {et~al.} 2013, \apjl,
  762, L21

\bibitem[{{MacGregor} {et~al.}(2017){MacGregor}, {Matr{\`a}}, {Kalas},
  {Wilner}, {Pan}, {Kennedy}, {Wyatt}, {Duchene}, {Hughes}, {Rieke}, {Clampin},
  {Fitzgerald}, {Graham}, {Holland}, {Pani{\'c}}, {Shannon}, \& {Su}}]{macg17}
{MacGregor}, M.~A., {Matr{\`a}}, L., {Kalas}, P., {et~al.} 2017, \apj, 842, 8

\bibitem[{{Matr{\`a}} {et~al.}(2019){Matr{\`a}}, {Wyatt}, {Wilner}, {Dent},
  {Marino}, {Kennedy}, \& {Milli}}]{matr19}
{Matr{\`a}}, L., {Wyatt}, M.~C., {Wilner}, D.~J., {et~al.} 2019, \aj, 157, 135

\bibitem[{{Matr{\`a}} {et~al.}(2017){Matr{\`a}}, {Dent}, {Wyatt}, {Kral},
  {Wilner}, {Pani{\'c}}, {Hughes}, {de Gregorio-Monsalvo}, {Hales}, {Augereau},
  {Greaves}, \& {Roberge}}]{matr17}
{Matr{\`a}}, L., {Dent}, W.~R.~F., {Wyatt}, M.~C., {et~al.} 2017, \mnras, 464,
  1415

\bibitem[{{Matthews} {et~al.}(2014){Matthews}, {Kennedy}, {Sibthorpe}, {Booth},
  {Wyatt}, {Broekhoven-Fiene}, {Macintosh}, \& {Marois}}]{matt14}
{Matthews}, B., {Kennedy}, G., {Sibthorpe}, B., {et~al.} 2014, \apj, 780, 97

\bibitem[{{Mayor} {et~al.}(2011){Mayor}, {Marmier}, {Lovis}, {Udry},
  {S{\'e}gransan}, {Pepe}, {Benz}, {Bertaux}, {Bouchy}, \& {Dumusque}}]{mayo11}
{Mayor}, M., {Marmier}, M., {Lovis}, C., {et~al.} 2011, arXiv e-prints,
  arXiv:1109.2497

\bibitem[{{Miley} {et~al.}(2018){Miley}, {Pani{\'c}}, {Wyatt}, \&
  {Kennedy}}]{mile18}
{Miley}, J.~M., {Pani{\'c}}, O., {Wyatt}, M., \& {Kennedy}, G.~M. 2018, \aap,
  615, L10

\bibitem[{{Morrison} \& {Malhotra}(2015)}]{morr15}
{Morrison}, S., \& {Malhotra}, R. 2015, \apj, 799, 41

\bibitem[{{Mouillet} {et~al.}(1997){Mouillet}, {Larwood}, {Papaloizou}, \&
  {Lagrange}}]{moui97}
{Mouillet}, D., {Larwood}, J.~D., {Papaloizou}, J.~C.~B., \& {Lagrange}, A.~M.
  1997, \mnras, 292, 896

\bibitem[{{Murray} \& {Dermott}(1999)}]{murr99}
{Murray}, C.~D., \& {Dermott}, S.~F. 1999, {Solar system dynamics}

\bibitem[{{Murray} \& {Holman}(1997)}]{murr97}
{Murray}, N., \& {Holman}, M. 1997, \aj, 114, 1246

\bibitem[{{Mustill} \& {Wyatt}(2011)}]{must11}
{Mustill}, A.~J., \& {Wyatt}, M.~C. 2011, \mnras, 413, 554

\bibitem[{{Petrovich}(2015)}]{petr15}
{Petrovich}, C. 2015, \apj, 808, 120

\bibitem[{{Quanz} {et~al.}(2015){Quanz}, {Crossfield}, {Meyer}, {Schmalzl}, \&
  {Held}}]{quan15}
{Quanz}, S.~P., {Crossfield}, I., {Meyer}, M.~R., {Schmalzl}, E., \& {Held}, J.
  2015, International Journal of Astrobiology, 14, 279

\bibitem[{{Quillen}(2006)}]{quil06}
{Quillen}, A.~C. 2006, \mnras, 372, L14

\bibitem[{{Quillen} {et~al.}(2007){Quillen}, {Morbidelli}, \& {Moore}}]{quil07}
{Quillen}, A.~C., {Morbidelli}, A., \& {Moore}, A. 2007, \mnras, 380, 1642

\bibitem[{{Reche} {et~al.}(2008){Reche}, {Beust}, {Augereau}, \&
  {Absil}}]{rech08}
{Reche}, R., {Beust}, H., {Augereau}, J.~C., \& {Absil}, O. 2008, \aap, 480,
  551

\bibitem[{{Rein} \& {Liu}(2012)}]{rein12}
{Rein}, H., \& {Liu}, S.-F. 2012, \aap, 537, A128

\bibitem[{{Rein} \& {Spiegel}(2015)}]{rein15a}
{Rein}, H., \& {Spiegel}, D.~S. 2015, \mnras, 446, 1424

\bibitem[{{Rein} \& {Tamayo}(2015)}]{rein15b}
{Rein}, H., \& {Tamayo}, D. 2015, \mnras, 452, 376

\bibitem[{{Ricci} {et~al.}(2015){Ricci}, {Carpenter}, {Fu}, {Hughes}, {Corder},
  \& {Isella}}]{ricc15}
{Ricci}, L., {Carpenter}, J.~M., {Fu}, B., {et~al.} 2015, \apj, 798, 124

\bibitem[{{Roberge} {et~al.}(2019){Roberge}, {Fischer}, \& {Peterson}}]{robe19}
{Roberge}, A., {Fischer}, D., \& {Peterson}, B. 2019, in \baas, Vol.~51, 199

\bibitem[{{Rodigas} {et~al.}(2014){Rodigas}, {Malhotra}, \& {Hinz}}]{rodi14}
{Rodigas}, T.~J., {Malhotra}, R., \& {Hinz}, P.~M. 2014, \apj, 780, 65

\bibitem[{{Schlichting}(2014)}]{schl14}
{Schlichting}, H.~E. 2014, \apjl, 795, L15

\bibitem[{{Shannon} {et~al.}(2015){Shannon}, {Mustill}, \& {Wyatt}}]{shan15}
{Shannon}, A., {Mustill}, A.~J., \& {Wyatt}, M. 2015, \mnras, 448, 684

\bibitem[{{Sibthorpe} {et~al.}(2018){Sibthorpe}, {Kennedy}, {Wyatt},
  {Lestrade}, {Greaves}, {Matthews}, \& {Duch{\^e}ne}}]{sibt18}
{Sibthorpe}, B., {Kennedy}, G.~M., {Wyatt}, M.~C., {et~al.} 2018, \mnras, 475,
  3046

\bibitem[{{Su} {et~al.}(2019){Su}, {Ballering}, {Ertel}, {Gaspar}, {Kennedy},
  {Leisawitz}, {MacGregor}, {Matthews}, {Moro-Martin}, {Rieke}, {White},
  {Wilner}, \& {Wyatt}}]{su19}
{Su}, K., {Ballering}, N., {Ertel}, S., {et~al.} 2019, arXiv e-prints,
  arXiv:1903.10616

\bibitem[{{Su} {et~al.}(2009){Su}, {Rieke}, {Stapelfeldt}, {Malhotra},
  {Bryden}, {Smith}, {Misselt}, {Moro-Martin}, \& {Williams}}]{su09}
{Su}, K.~Y.~L., {Rieke}, G.~H., {Stapelfeldt}, K.~R., {et~al.} 2009, \apj, 705,
  314

\bibitem[{{Tamayo} {et~al.}(2019){Tamayo}, {Rein}, {Shi}, \& {Hernand
  ez}}]{tama19}
{Tamayo}, D., {Rein}, H., {Shi}, P., \& {Hernand ez}, D.~M. 2019, arXiv
  e-prints, arXiv:1908.05634

\bibitem[{{Telesco} {et~al.}(2000){Telesco}, {Fisher}, {Pi{\~n}a}, {Knacke},
  {Dermott}, {Wyatt}, {Grogan}, {Holmes}, {Ghez}, {Prato}, {Hartmann}, \&
  {Jayawardhana}}]{tele00}
{Telesco}, C.~M., {Fisher}, R.~S., {Pi{\~n}a}, R.~K., {et~al.} 2000, \apj, 530,
  329

\bibitem[{{Th{\'e}bault}(2009)}]{theb09}
{Th{\'e}bault}, P. 2009, \aap, 505, 1269

\bibitem[{{Th{\'e}bault} \& {Augereau}(2007)}]{theb07}
{Th{\'e}bault}, P., \& {Augereau}, J.~C. 2007, \aap, 472, 169

\bibitem[{{van der Walt} {et~al.}(2011){van der Walt}, {Colbert}, \&
  {Varoquaux}}]{vand11}
{van der Walt}, S., {Colbert}, S.~C., \& {Varoquaux}, G. 2011, Computing in
  Science and Engineering, 13, 22

\bibitem[{{Wilner} {et~al.}(2018){Wilner}, {MacGregor}, {Andrews}, {Hughes},
  {Matthews}, \& {Su}}]{wiln18}
{Wilner}, D.~J., {MacGregor}, M.~A., {Andrews}, S.~M., {et~al.} 2018, \apj,
  855, 56

\bibitem[{{Winn} \& {Fabrycky}(2015)}]{winn15}
{Winn}, J.~N., \& {Fabrycky}, D.~C. 2015, \araa, 53, 409

\bibitem[{{Wisdom}(1980)}]{wisd80}
{Wisdom}, J. 1980, \aj, 85, 1122

\bibitem[{{Wyatt}(2003)}]{wyat03}
{Wyatt}, M.~C. 2003, \apj, 598, 1321

\bibitem[{{Wyatt}(2008)}]{wyat08}
---. 2008, Annual Review of Astronomy and Astrophysics, 46, 339

\bibitem[{{Wyatt}(2018)}]{wyat18}
---. 2018, {Debris Disks: Probing Planet Formation}, 146

\bibitem[{{Wyatt} {et~al.}(1999){Wyatt}, {Dermott}, {Telesco}, {Fisher},
  {Grogan}, {Holmes}, \& {Pi{\~n}a}}]{wyat99}
{Wyatt}, M.~C., {Dermott}, S.~F., {Telesco}, C.~M., {et~al.} 1999, \apj, 527,
  918

\bibitem[{{Yelverton} \& {Kennedy}(2018)}]{yelv18}
{Yelverton}, B., \& {Kennedy}, G.~M. 2018, \mnras, 479, 2673

\end{thebibliography}

\appendix
\section{Simulation Setup for Figure 1}
\label{setup}
\begin{deluxetable}{cccccccccl}[htb!]
\tablecaption{Setup parameters for $N$-body simulations in Figure 1\label{tab:1}. Column 2-5 illustrates planet b's mass, semi-major axis, inclination, and eccentricity, respectively. Column 6-7 presents the semi-major axis range and the total number of test particles. Integrator types for the simulations are indicated in column 8.}
\tablehead{
\colhead{Debris Disk Feature} & \colhead{$m_{\textrm{b}}$} & \colhead{$a_{\textrm{b}}$} & \colhead{$i_{\textrm{b}}$} & \colhead{$e_{\textrm{b}}$} & \colhead{$a_{\textrm{par}}$} & \colhead{$N_{\textrm{tot}}$} & \colhead{Integration Time} & \colhead{Integrator} & \colhead{Notes} \\ 
\colhead{} & \colhead{(M$_{\textrm{Jup}}$)} & \colhead{(au)} & \colhead{(\degree)} & \colhead{} & \colhead{(au)} & \colhead{} & \colhead{(Myr)} & \colhead{} &\colhead{}
}
\startdata
Warp & 10 & 10 & 10 & 0 & 20--200 & 1e4 & 10 & whfast & edge-on view\\
Spiral arm & 10 & 10 & 10 & 0 & 20--200 & 1e4 & 10 & whfast & view at 30\degree-inc, 150\degree-pos angle\\
Offset & 1 & 10 & 0 & 0.4 & 20--200 & 1e4 & 10 & whfast &...\\
Azimuthal asymmetry & 1 & 100 & 0 & 0 & 20--200 & 7.5e3 & 10 & ias15 & $a_{\textrm{b}}$ decaying as $e^{-t/(20 \textrm{ Myr})}$\\
Edge \& gap & 1 & 16 & 0 & 0 & 16--26 & 4e3 & 10 & ias15 & ...\\
Radially thickened ring & 1 & 10 & 0 & 0 & 13 & 1e4 & 0.3 & whfast & ...\\
Scale height & 1 & 10 & 10 & 0 & 9--11 & 1e4 & 0.3 & whfast & ...\\
\enddata
\tablecomments{(a) $\Omega_{\textrm{b}}$ and $\varpi_{\textrm{b}}$ are fixed to be zero in all simulations. (b) All test particles are massless. (c) We assume a solar-mass host star. (d) $N$-body simulations run on REBOUND \citep{rein15a,rein15b} and REBOUNDx \citep{tama19}.}
\end{deluxetable}  

\section{Secular Perturbation Code}
\label{secular code}
The secular perturbation code is based on the Laplace-Lagrange approximation, a first-order approximation in $e$ and $i$ of the secular terms in the disturbing function which works appropriately for planets with low inclinations ($i\lesssim20\degree$) and eccentricities ($e\lesssim0.3$) (e.g., \citealt{murr99}, Chapter 7). We describe a planetesimal's inclination and longitude of ascending node using its inclination vector $(p, q)$ as $i = \sqrt{p^2+q^2}$, $\Omega = \tan^{-1}{(p/q)}$ and a planetesimal's eccentricity and longitude of pericenter using its eccentricity vector $(h, k)$ as $e = \sqrt{h^2+k^2}$, $\varpi = \tan^{-1}{(h/k)}$. The inclination vector $(p, q)$ and the eccentricity vector $(h, k)$ have the following form:
\begin{equation}
    \begin{split}
        p &= i_{\text{free}}\sin{(Bt + \gamma)}+p_{\textrm{forced}}(t), \\ 
        q &= i_{\text{free}}\cos{(Bt+\gamma)}+q_{\textrm{forced}}(t), \\
        h &= e_{\text{free}}\sin{(At + \beta)}+h_{\textrm{forced}}(t),\\
        k &= e_{\text{free}}\cos{(At + \beta)}+k_{\textrm{forced}}(t).\\
    \end{split}
\end{equation}
The first term in both the inclination and eccentricity vector is the free element. The free inclination $i_{\text{free}}$, the phase offset $\gamma$, the free eccentricity $e_{\text{free}}$, and the phase offset $\beta$ depend on the initial condition of the planetesimal. Unless stated otherwise, we set initial orbital parameters of a planetesimal to 0 except its semi-major axis. The nodal regression rate $B$ and the pericentre precession rate $A$ for a planetesimal located at the semi-major axis $a$ have a similar form:
\begin{equation} \label{eqn:precession}
    \begin{gathered}
    B = -A = -n\frac{1}{4}\sum_{j=1}^{N_{\textrm{plt}}}\frac{m_j}{m_*}\alpha_j\bar{\alpha}_jb_{3/2}^{(1)}(\alpha_j),     
  \end{gathered}
\end{equation}
where $n$ is the orbital frequency of the planetesimal, $N_{\textrm{plt}}$ is the total number of planets in the system, $m_*$ is the stellar mass, $\alpha_j$ and $\bar{\alpha}_j$ for $j$-th planet at the semi-major axis of $a_j$ follow the formula:
\begin{equation} \label{eqn: alpha_j}
        \alpha_j \text{, } \bar{\alpha}_j = 
            \begin{dcases}
                a_j/a \text{, } 1 &\text{\quad if \quad} a_j < a \\
                a/a_j &\text{\quad if \quad} a_j > a,
            \end{dcases}
\end{equation}
and $b_{3/2}^{(1)}$ is the Laplace coefficient. The second term in the inclination/eccentricity vector is the forced element, ($p_{\textrm{forced}}$, $q_{\textrm{forced}}$) and ($h_{\textrm{forced}}$, $k_{\textrm{forced}}$), reflecting the secular perturbation of a planetesimal from planets:
\begin{equation} \label{eqn:forced}
    \begin{split}
        p_{\textrm{forced}} &= \sum_{i=1}^{N_{\textrm{plt}}}\frac{-\mu_i}{B-f_i}\sin{(f_it+\gamma_i)}, \\
        q_{\textrm{forced}} &= \sum_{i=1}^{N_{\textrm{plt}}}\frac{-\mu_i}{B-f_i}\cos{(f_it+\gamma_i)}, \\
        h_{\textrm{forced}} &= \sum_{i=1}^{N_{\textrm{plt}}}\frac{-\nu_i}{A-g_i}\sin{(g_it+\beta_i)}, \\
        k_{\textrm{forced}} &= \sum_{i=1}^{N_{\textrm{plt}}}\frac{-\nu_i}{A-g_i}\cos{(g_it+\beta_i)},
    \end{split}    
\end{equation}
where $\mu_i$, $f_i$, $\gamma_i$, $\nu_i$, $g_i$, and $\beta_i$ are the $i$-th mode of a $N_{\textrm{plt}}$-planet system. $\gamma_i$ and $\beta_i$ can be determined from the initial condition of the planets. $f_i$ and $g_i$ are the $i$-th eigenvalue of matrices $B$ and $A$, written as
\begin{equation} \label{eqn:matrix}
    \begin{gathered}
        B_{jk} = +n_j\frac{1}{4}\frac{m_k}{m_*+m_j}\alpha_{jk}\bar{\alpha}_{jk}b_{3/2}^{(1)}(\alpha_{jk}), \
        B_{jj} = -B_{jk}; \\
        A_{jk} = -n_j\frac{1}{4}\frac{m_k}{m_*+m_j}\alpha_{jk}\bar{\alpha}_{jk}b_{3/2}^{(2)}(\alpha_{jk}), \
        A_{jj} = -B_{jj},
    \end{gathered}
\end{equation}
where $j \neq k$, $n_j$ is the orbital frequency of the $j$-th planet, $m_k$ and $m_j$ are the mass of the $k$-th and $j$-th planet. $\alpha_{jk}$ and $\bar{\alpha}_{jk}$ in the equation are unitless numbers written as
\begin{equation}
        \alpha_{jk} \text{, } \bar{\alpha}_{jk} = 
            \begin{dcases}
                a_k/a_j \text{, } 1 &\text{\quad if \quad} a_j > a_k \\
                a_j/a_k &\text{\quad if \quad} a_j < a_k,
            \end{dcases}
\end{equation}
and $b_{3/2}^{(1)}$ and $b_{3/2}^{(2)}$ are the Laplace coefficients written as
\begin{equation} \label{eqn:coeff}
    \begin{gathered}
        b_{3/2}^{(1)}(\alpha) = \frac{1}{\pi}\int_{0}^{2\pi} \frac{\cos \psi d\psi}{(1-2\alpha\cos\psi+\alpha^2)^{3/2}}\\
        b_{3/2}^{(2)}(\alpha) = \frac{1}{\pi}\int_{0}^{2\pi} \frac{\cos2\psi d\psi}{(1-2\alpha\cos\psi+\alpha^2)^{3/2}}
    \end{gathered}
\end{equation}
$\mu_i$ and $\nu_i$ in Equation~(\ref{eqn:forced}) can be expressed as
\begin{equation} \label{eqn:nu}
    \begin{gathered}
        \mu_i = \sum_{j=1}^{N_{\textrm{plt}}}B_jI_{ji}, \
        \nu_i = \sum_{j=1}^{N_{\textrm{plt}}}A_je_{ji},
    \end{gathered}
\end{equation}
where $I_{ji}$ and $e_{ji}$ are the $i$-th eigenvector of matrices described in Equation~(\ref{eqn:matrix}) normalized by the initial conditions of planets, and
\begin{equation}\label{eqn:Aj}
    \begin{gathered}
    B_j = +n\frac{1}{4}\frac{m_j}{m_*}\alpha_j\bar{\alpha}_jb_{3/2}^{(1)}(\alpha_j), \
    A_j = -n\frac{1}{4}\frac{m_j}{m_*}\alpha_j\bar{\alpha}_jb_{3/2}^{(2)}(\alpha_j).
  \end{gathered}
\end{equation}
$\alpha_j$ here has the similar expression as Equation~(\ref{eqn: alpha_j}). Given the secular code written from the analytical solution above, we are able to explore a full range of configurations of a multi-planet model and study the planets' effects on the disk. 
\end{document}